\PassOptionsToPackage{comma,authoryear,compress}{natbib}
\documentclass[11pt,letterpaper]{mystyle}

\usepackage[all]{hypcap}
\definecolor{lightblue}{rgb}{0.22,0.45,0.70}
\hypersetup{
    colorlinks=true,
    citecolor=lightblue,
}
\providecommand{\Description}[1]{}
\usepackage{array}
\newcolumntype{L}[1]{>{\raggedright\arraybackslash}p{#1}}
\usepackage{mathtools}

\usepackage{multirow}
\usepackage{graphicx}

\renewcommand{\mathbf}{\boldsymbol}
\newcolumntype{Y}{>{\raggedright\arraybackslash}X}

\title{Uni-XAS: Alignment-Driven Bidirectional Multimodal Learning for X-ray Absorption Spectroscopy}

\runningtitle{Proceedings of the 34th ACM International Conference on Multimedia (MM '26), November 10--14, 2026, Rio de Janeiro, Brazil}

\author[1,\textdagger]{Suyang Zhong}
\author[1,\textdagger]{Yuhao Zhao}
\author[1]{Boying Huang}
\author[1]{Fanjie Xu}
\author[1]{Pengwei Xu}
\author[2]{Haoyi Tao}
\author[2,*]{Xi Fang}
\author[1,*]{Jun Cheng}
\author[1,*]{Fujie Tang}

\affil[1]{State Key Laboratory of Physical Chemistry of Solid Surfaces, iChEM, IKKEM, School of Intelligent Manufacturing, Institute of Artificial Intelligence, College of Chemistry and Chemical Engineering, Xiamen University, Xiamen 361005, China}
\affil[2]{DP Technology, Shanghai 200030, China}

\authornotes{%
  \textsuperscript{\textdagger}\,These authors contributed equally to this work.\par
  \textsuperscript{*}\,Corresponding authors to whom correspondence should be addressed.%
}

\begin{document}

\begin{abstract}
X-ray absorption spectroscopy (XAS) is a key technique for probing local atomic environments, yet learning-based modeling must bridge two heterogeneous modalities: 1D continuous spectra and 3D atomic structures. Existing approaches typically decouple forward spectrum prediction and inverse structure inference into separate regression tasks, hindering shared representation learning. Moreover, severe permutation ambiguity among identical atoms often limits inverse modeling to coarse structure descriptors rather than explicit 3D structure generation. In this work, we present Uni-XAS, a unified benchmark and learning framework that reframes bidirectional XAS modeling as a cross-modal alignment and conditional generation problem. We first propose XASLip, an alignment recipe coupling a physics-aware spectral encoder with an absorber-aware manifold optimization strategy to resolve fine-grained intra-element coordination variations. Building upon this shared latent space, we formulate forward prediction as anchored absolute-spectrum generation via retrieval-augmented decoding, effectively preventing physical scale collapse and energy drift. For the inherently ill-posed inverse problem, we introduce Permutation-Rectified Flow Matching, which integrates type-wise optimal transport into a continuous generative flow to provide a principled solution to ligand permutation ambiguity without relying on heavy high-order equivariant architectures. Evaluated on a large-scale standardized benchmark of 328,839 structure--spectrum pairs, Uni-XAS demonstrates strong performance in cross-modal retrieval, accurate absolute-spectrum prediction, and composition-conditional 3D structure generation, establishing a scalable, reproducible, and protocol-consistent foundation for multimodal learning and standardized evaluation in scientific spectroscopy.

\includegraphics[height=1em]{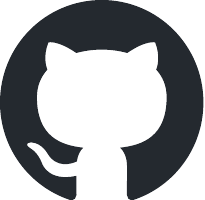}
\textbf{Code Repository}: \url{https://github.com/zsy-suyang/Uni-XAS}

\textbf{Contact}: {tangfujie@xmu.edu.cn}

\end{abstract}


\maketitle

\section{Introduction}
    X-ray absorption spectroscopy (XAS) is a crucial analytical technique for probing local atomic environments~\cite{ravel2010simultaneous,kerr2022characterization,cutsail2022challenges}. By analyzing X-ray absorption near-edge structure, known as XANES, researchers can decode complex local coordination geometries and electronic states~\cite{liu2021dynamic,kang2023cr,carbone2023lightshow}. Accelerating XAS analysis through machine learning has become a prominent direction in AI for Science~\cite{liang2023decoding,Deyu2023Carbon,chen2024robust,protsenko2025fingerprint}, yet it remains limited by the modality gap.

    Fundamentally, XAS modeling involves mapping between two structurally distinct scientific modalities: 1D continuous spectral sequences and absorber-centered 3D atomic structures. Because XANES is governed primarily by the immediate scattering environment, we explicitly scope this structural modality to the local coordination geometry around the absorbing atom, rather than the infinite periodic crystal lattice. Unlike conventional vision and language, aligning these geometric and spectral signals poses unique challenges beyond existing multimodal benchmarks. 
    
\begin{figure}[t]
        \centering
        \includegraphics[width=0.9\linewidth]{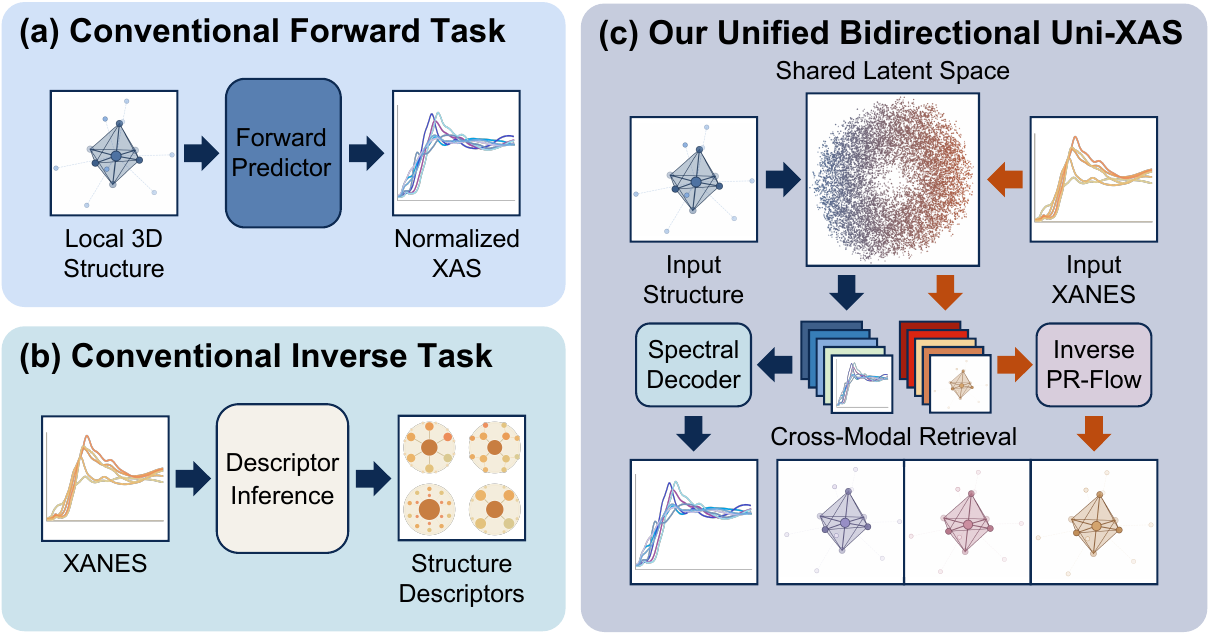}
        \caption{Comparison of XAS modeling pipelines. (a) Forward surrogate prediction. (b) Descriptor-level inverse interpretation. (c) Uni-XAS establishes a shared latent interface supporting cross-modal retrieval, retrieval-augmented forward prediction, and conditional 3D local-structure generation.}
        \Description{A block diagram showing three different XAS modeling pipelines: forward-only, inverse interpretation, and the proposed Uni-XAS shared latent interface.}
        \label{fig:pipeline}
    \end{figure}

    As illustrated in Figure~\ref{fig:pipeline}, existing learning-based approaches treat the structure-to-spectrum forward prediction~\cite{kharel2025omnixas,wang2025xastruct,lin2026cgxas} and the spectrum-to-structure inverse interpretation~\cite{na2025interpretable,zhong2026solvinginverseproblemxray} as separate regression problems. This decoupled paradigm suffers from three critical limitations: First, it hinders the formation of a shared cross-modal representation, making cross-directional knowledge transfer impossible. Second, forward surrogates predominantly regress normalized spectral shapes~\cite{zhan2025graph,kharel2025omnixas,wang2025xastruct}, discarding critical physical scales, such as absolute energy anchors. Third, applying deterministic regressors to the inherently one-to-many inverse mapping causes structural collapse, reducing most inverse studies to predicting coarse 1D descriptors such as oxidation states and coordination numbers~\cite{wang2025xastruct} rather than explicit 3D atomic coordinates. Compounding these algorithmic bottlenecks, the field lacks a large-scale, protocol-consistent benchmark and standardized reference suite. Prior studies are often reported on task-specific, non-comparable, and sometimes non-public settings, making fair cross-paper assessment difficult and impeding the development of generalized cross-modal foundation models. Motivated by cross-modal alignment methods~\cite{radford2021learning,zhai2023sigmoid} such as SigLIP, we treat XAS as a multimodal matching problem between absorber-centered 3D structures and 1D spectra. To address these challenges, we propose Uni-XAS, a unified benchmark and bidirectional framework which learns a shared latent interface that is reused for cross-modal retrieval, anchored absolute-spectrum prediction through Retrieval-Augmented Generation, and composition-conditional local-coordinate generation.
    
    Building such a unified multimodal system for XAS requires addressing several domain-specific challenges. First, fine-grained intra-element discrimination is essential. In standard multimodal retrieval, separating distinct classes is straightforward~\cite{radford2021learning,zhai2023sigmoid,tschannen2025siglip}. In contrast, while distinct absorbing elements in XAS exhibit macroscopic differences in baseline absorption thresholds, spectra originating from the same element differ only in subtle near-edge morphological variations. Therefore, cross-modal alignment must be sensitive to fine-grained coordination perturbations rather than broad compositional differences. Second, forward modeling must preserve absolute physical scales. Unlike arbitrary temporal sequences, XANES spectra possess an absolute energy threshold anchor, denoted as $E_0$, and non-normalized intensity scales. Forward prediction must strictly adhere to these physical constraints rather than relying on standard shape-matching objectives. Third, the framework must handle the one-to-many mapping inherent in 3D generation. Because identical atoms in the surrounding chemical environment, such as multiple oxygen atoms in a coordination shell, are physically indistinguishable, their arbitrary ordering introduces severe permutation ambiguity, akin to the challenge of unordered 3D point cloud generation in computer vision. Standard generative models struggle with this, often producing collapsed or physically unviable geometries due to conflicting training gradients.

    To address these challenges and provide a benchmark testbed, we create a standardized benchmark comprising 328,839 high-fidelity structure--spectrum pairs. Built upon this data engine, we design Uni-XAS through a progressive learning strategy. We first establish the shared latent space via XASLip, an alignment recipe that couples physics-aware spectral encoding with an absorber-aware manifold optimization strategy. This forces the latent space to focus on fine-grained, intra-element structural variations rather than macroscopic elemental separation. We then freeze this interface and deploy it for downstream modeling. This shared representation unifies both directions, overcoming the decoupled regression paradigms of prior works. For forward prediction, we formulate the task as anchored absolute-spectrum generation. To reduce physically implausible deviations, we introduce a retrieval-augmented decoding mechanism that leverages the shared space to calibrate the parametric prediction using verified spectral priors.
    
    For the challenging inverse task, we move beyond traditional descriptor-level inference by formulating it as conditional 3D local-structure generation. To explicitly address the aforementioned permutation ambiguity, we propose Permutation-Rectified Flow Matching (PR-Flow). By integrating a type-wise optimal transport assignment among identical atom types before flow supervision, PR-Flow establishes permutation-consistent training targets. Rather than relying on computationally heavy high-order equivariant parameterizations, our inverse decoder operates on mean-centered coordinates and uses lightweight $E(3)$-aware coordinate updates constructed from pairwise distances and relative displacements. Coupling this permutation-rectified objective with retrieval-augmented structural conditioning from the frozen latent space yields a stable and geometry-consistent generative process at benchmark scale. While bridging the sim-to-real gap for experimental spectra remains an open challenge, Uni-XAS provides a rigorously benchmarked foundation for applying modern multimodal representation learning to scientific spectroscopy.
    
    Our main contributions are summarized as follows:
    \begin{itemize}[leftmargin=*, nosep]
    \item \textbf{Unified Benchmark and Evaluation:} We construct a standardized, large-scale benchmark of 328,839 structure--spectrum pairs. This provides the first unified evaluation suite for cross-modal retrieval, absolute-spectrum prediction, and inverse composition-conditional 3D structure generation under a shared data contract.
    \item \textbf{XASLip Alignment \& RAG Forward Prediction:} We propose XASLip, coupling a physics-aware spectral encoder with an absorber-aware manifold optimization strategy. Leveraging this shared interface, we formulate forward modeling as anchored absolute-spectrum prediction via retrieval-augmented decoding, preserving physical scales and preventing energy drift.
    \item \textbf{Geometry-Consistent and Efficient Inverse Generation via PR-Flow:} To address the ill-posed inverse problem, we introduce PR-Flow. Integrating type-wise optimal transport into a continuous generative flow resolves ligand permutation ambiguity while coupling centered-coordinate training with lightweight decoder-level $E(3)$-equivariant coordinate updates, establishing an efficient baseline for composition-conditioned 3D generation.
    \end{itemize}

\section{Related Work}
    \subsection{Learning-based XAS Modeling}
    Classical XAS interpretation relies on curated reference matching or computationally expensive \textit{ab initio} calculations, such as FEFF~\cite{kas2021advanced,kerr2022characterization}, which use multiple-scattering and Green’s function formalisms. To bypass these computational bottlenecks, machine learning-based surrogates have transitioned from shallow regressors using handcrafted descriptors to Graph Neural Networks and Transformers~\cite{muller2016quick,zhan2025graph}. Despite their efficiency, existing methods are conceptually fragmented. First, they are limited to forward prediction. Second, they predominantly predict normalized spectral shapes, discarding the physical absorption scale and the critical absolute energy anchor, leading to systemic energy drift. Crucially, by treating XAS prediction as an isolated unimodal regression task, they do not explicitly construct a shared cross-modal representation space. This fragmentation prevents the realization of bidirectional consistency, cross-modal retrieval, and shared latent knowledge transfer. In contrast, Uni-XAS treats bidirectional XAS modeling fundamentally as a unified cross-modal alignment problem.

    \subsection{Multimodal Alignment and Reference Matching}

    Contrastive alignment paradigms~\cite{radford2021learning,yao2021filip,zhai2023sigmoid,li2022grounded}, such as CLIP and SigLIP, have significantly advanced multimodal representation learning, inspiring unified cross-modal spaces in AI for Science~\cite{liu2023multi}. Historically, classical XAS analysis heavily relied on traditional reference matching or library searching algorithms~\cite{zhu2021emerging,virga2023structural,kwon2024spectroscopy} to identify structures from spectral databases. However, directly transferring modern large-scale alignment techniques to replace these classical methods introduces a distinctive domain challenge.
    
    Unlike conventional multimodal settings where inter-class variation dominates, XAS requires distinguishing subtle intra-element variations induced by minute shifts in local coordination geometry. Under standard batch-based contrastive objectives, these fine-grained distinctions often collapse, or the optimization suffers from hard-negative explosion. To overcome this, we introduce XASLip, an alignment recipe specifically designed for fine-grained structure--spectrum matching. By coupling physics-aware spectral encoding with an absorber-aware manifold optimization strategy, XASLip explicitly pushes cross-modal alignment from coarse elemental separation toward coordination-level discrimination, offering a learned, scalable alternative to traditional library matching.
    
    \subsection{Retrieval-Augmented Structure Generation}
    Progressing from spectrum to coordinate generation is fundamentally challenged by the ill-posed one-to-many 1D-to-3D mapping, further exacerbated by permutation ambiguity among chemically identical ligands in local environments. Prior works in 3D vision and molecular modeling have explored permutation-invariant generation for point clouds and molecules~\cite{luo2021diffusion,hoogeboom2022equivariant,klein2023equivariant}. However, directly adapting these approaches to XAS is non-trivial, as the generative process must be conditioned on highly multiplexed 1D physical signals, where fully equivariant architectures often incur substantial computational overhead at scale. Retrieval-Augmented Generation (RAG), originally developed in natural language processing, has recently been extended to multimodal and geometric domains to improve generation fidelity via external memory or reference retrieval. In the context of 3D structure modeling, retrieval mechanisms can provide target-aware priors or initialization cues~\cite{seo2024retrieval,huang2024interaction,lee2024molecule}, which are particularly beneficial under ill-posed inverse settings.
    
    In XAS and near-edge absorption spectroscopy, inverse modeling traditionally focuses on descriptor regression, reference matching, or fitting-based recovery of limited structural parameters, rather than explicit 3D coordinate generation under spectrum conditioning~\cite{wang2023interpretable,kwon2024spectroscopy,wang2025xastruct}. Concurrent work explores direct spectrum-to-structure generation via equivariant diffusion, but remains strictly confined to site-specific coordination recovery within specific material families such as silicon-oxygen, precluding generalization across diverse elements~\cite{okubo2026generative}. Despite these advances, a large-scale, protocol-consistent benchmark evaluating retrieval alignment, forward prediction, and inverse 3D generation remains absent, limiting systematic progress in unified cross-modal XAS modeling.
    
\begin{figure*}[t]
    \centering
    \includegraphics[width=0.98\linewidth]{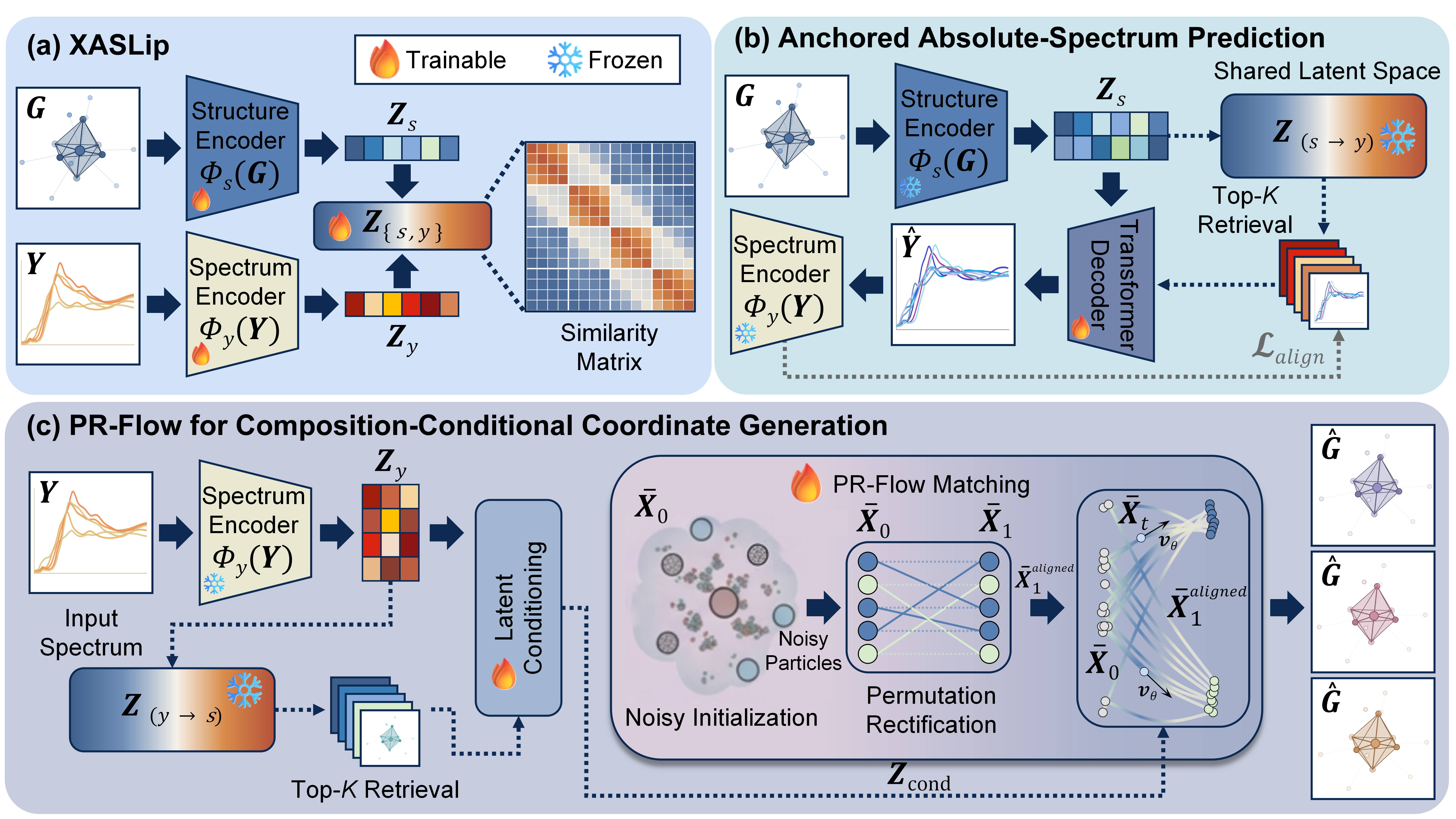}
    \caption{Uni-XAS architecture based on a Shared Latent Space ($\mathcal{Z}$). \textbf{(a) XASLip Alignment:} Encoders map 3D structures and 1D spectra to shared embeddings ($\mathbf{Z}_s, \mathbf{Z}_y$) via an absorber-aware objective. \textbf{(b) Absolute-Spectrum Prediction:} Query $\mathbf{Z}_s$ retrieves spectral priors ($\mathbf{Z}$) from the frozen space to augment decoding, regularized by a Latent Consistency Constraint ($\mathcal{L}_{\text{align}}$). \textbf{(c) PR-Flow Inverse Generation:} Type-wise optimal transport assigns initial noise ($\mathbf{X}_0$) to identical atoms to resolve permutation ambiguity. Flow trajectory ($\mathbf{X}_t$) is optimized toward the aligned target ($\mathbf{X}_1^{\text{aligned}}$) under retrieved structural conditioning.}
    \Description{A detailed three-panel architectural schematic of the Uni-XAS framework. Panel (a) displays two parallel encoder blocks taking a 3D atomic structure and a 1D spectrum as inputs, with arrows pointing toward a shared latent embedding space. Panel (b) shows a structural query fetching retrieved spectral priors from a frozen memory bank block, which then feed into a decoding module to output a predicted spectrum, linked to a latent consistency loss component. Panel (c) visualizes the inverse generation process, showing an initial noise distribution mapped to specific atomic nodes via an optimal transport block, followed by arrows representing a continuous flow matching trajectory that targets a 3D aligned structure, conditioned on structural prototypes.}
    \label{fig:model}
\end{figure*}

\section{Data Engine: Paired Structure--Spectrum Benchmark}\label{sec:data}

    To bridge the semantic and geometric gap between 3D structures and 1D spectra, we develop a scalable pipeline that mines and standardizes paired observations from the XASDataLibrary~\cite{xasdatalibrary} and Materials Project~\cite{jain2013commentary}. Spectra are linked to their crystallographic sources via deterministic metadata filtering, which is detailed in Section B of the Supplementary Information (SI). We shift the structural modality from infinite periodic crystals to physically relevant, absorber-centric local 3D graphs bounded by a predefined distance cutoff. For the spectral modality, we preserve the non-normalized absolute intensity and extract the physical energy threshold anchor $E_0$. For learning, the anchor is mapped into an absorber-specific normalized coordinate $\tilde{E}_0\in[0,1]$, yielding a unified 129-dimensional training target $\mathbf{Y}=[\tilde{E}_0,\mathbf{I}]$. All reported anchor errors are converted back to physical electronvolts at evaluation time.
    
    This yields a protocol-consistent benchmark of 328,839 high-fidelity pairs across 43 absorber atomic elements. To ensure a balanced evaluation, we employ an 80/20 train/test split stratified by absorber species. This large-scale curation effectively highlights realistic multimodal challenges—aligning long-tailed categories and heterogeneous local neighborhoods—supporting robust evaluation across all proposed tasks.
    
\section{Methods}\label{sec:methods}

    \subsection{Unified Bidirectional Framework}
    
    We adopt a staged, interface-unified learning paradigm bridging the 3D structure domain $\mathcal{S}$ and the standardized spectral domain $\mathcal{Y}$. Each paired sample is defined as:
    \begin{equation}\label{eq:sample_def}
        (\mathbf{G}, \mathbf{Y}), \quad \text{with} \quad \mathbf{G}=(\mathbf{V},\mathbf{X}), \quad \mathbf{Y}=[\tilde{E}_0, \mathbf{I}],
    \end{equation}
    where the absorber-centered local graph $\mathbf{G} \in \mathcal{S}$ comprises node features $\mathbf{V}$ and 3D spatial coordinates $\mathbf{X}\in\mathbb{R}^{N\times 3}$. The physical edge anchor $E_0$ is linearly normalized into an intra-element coordinate $\tilde{E}_0 \in [0,1]$ using precomputed elemental bounds to avoid severe numerical imbalance during optimization, whereas reported anchor metrics are computed after denormalization.
    
    Two dedicated encoders map these modalities into a shared $\ell_2$-normalized latent space $\mathcal{Z}$:
    \begin{equation}\label{eq:encoders}
        \mathbf{z}_s = \Phi_s(\mathbf{G}), \qquad \mathbf{z}_y = \Phi_y(\mathbf{Y}), \quad \text{where} \quad \mathbf{z}_s, \mathbf{z}_y \in \mathcal{Z}.
    \end{equation}
    For the structure branch, $\Phi_s$ is instantiated as a 3D geometric Transformer. By projecting continuous pairwise distances into Gaussian basis functions and injecting them into the self-attention mechanism, it establishes a robust $SE(3)$-invariant baseline to capture coordination-relevant structure (see SI Section D for details).
    
    To ensure optimization stability and prevent cross-task gradient interference, we adopt a progressive three-stage training protocol (Figure~\ref{fig:model}): (1) joint optimization of the dual encoders to construct the alignment space; (2) training the forward absolute-spectrum predictor using the frozen space for spectral retrieval; and (3) training the inverse PR-Flow generator conditioned on retrieved structural prototypes. This ensures modular reuse of the aligned latent interface and clear attribution of stage-wise contributions.
    
    \subsection{Physics-Aware Spectral Encoder}
    Unlike generic sequential signals, XAS carries explicit physical structure: the intensity profile $\mathbf{I}$ reflects the absorption cross-section, while the near-edge region contains coordination-sensitive resonances. To capture this, we construct a multi-channel sequence combining raw intensity with its first and second numerical derivatives, computed after Gaussian smoothing to suppress noise. This design is physically motivated: the first derivative isolates the steepest rise to robustly estimate the absorption onset, while the second derivative highlights local curvature associated with pre-edge shoulders and white lines. Prepended with the normalized energy anchor token $\tilde{E}_0$, this sequence allows the model to jointly capture absolute physical scale and spectral morphology. The tokens are then processed through $L$ stacked blocks combining 1D depth-wise separable convolution with multi-head self-attention.
    
    To explicitly localize this near-edge region, we introduce an edge-onset-guided pooling mechanism. We define the onset index $t^{\star}$ as the maximizer of the smoothed first derivative. Anchoring a symmetric window at $t^{\star}$, we pool the hidden states to extract an edge-localized spectral summary $\tilde{\mathbf{z}}_y$. A global embedding $\mathbf{z}_y$, capturing the broad absorption envelope, is simultaneously obtained via full-sequence average pooling. Finally, unpooled token features are output to support downstream fine-grained alignment. This dual-resolution design uses interpretable localization to constrain local spectral aggregation, injecting essential domain priors without architectural overhead to directly facilitate fine-grained, intra-element cross-modal matching (details in SI Section D).
        
    \subsection{XASLip: Cross-Modal Alignment}
    Conventional cross-modal objectives prioritize inter-class separation. However, transferring them to XAS is insufficient because the fundamental discriminative bottleneck lies in resolving fine-grained coordination variations among samples sharing the identical absorbing element. To address this, XASLip introduces an XAS-specific alignment method combining physics-aware spectral tokenization with an absorber-aware manifold optimization strategy.
    
    \paragraph{Dual-Stream Representation.}
    Internally, each encoder $\Phi_m$ ($m \in \{s, y\}$) processes its input to produce a global feature $\mathbf{g}_m$ and a sequence of token-level features $\mathbf{T}_m$. To establish the dual-stream representation, the global branch directly projects $\mathbf{g}_m$ into the shared space, while the fine-grained branch dynamically aggregates $\mathbf{T}_m$ using learnable Multi-Head Attention Pooling (MAP):
    \begin{equation}
        \mathbf{z}_m = \mathrm{Proj}_g(\mathbf{g}_m), \qquad \tilde{\mathbf{z}}_m = \mathrm{Proj}_l\big(\mathrm{MAP}(\mathbf{T}_m)\big).
    \end{equation}
    Here, $\mathrm{Proj}_{g,l}(\cdot)$ are linear heads mapping to the $\ell_2$-normalized latent space. By applying query-based attention, this design maintains a stable global teacher stream while extracting coordination-sensitive motifs for the fine-grained student stream.
    
    \paragraph{Hierarchical Alignment via Matching, Consistency, and Absorber Regularization.}
    We use an uncoupled pairwise sigmoid objective to enforce instance-level cross-modal matching in both the global and fine-grained branches. For the batch size $B$:
    \begin{equation}
        \begin{aligned}
            \mathcal{L}_{\mathrm{global}} &= -\frac{1}{B^2}\sum_{i,j} \log \sigma \Big( s_{ij}\big(\alpha\,\mathbf{z}_{s,i}^{\top}\mathbf{z}_{y,j}+\beta\big) \Big), \\
            \mathcal{L}_{\mathrm{fg}} &= -\frac{1}{B^2}\sum_{i,j} \log \sigma \Big( s_{ij}\big(\alpha\,\tilde{\mathbf{z}}_{s,i}^{\top}\tilde{\mathbf{z}}_{y,j}+\beta\big) \Big),
        \end{aligned}
    \end{equation}
    where $s_{ij}=1$ for exact pairs and $s_{ij}=-1$ otherwise, and $\alpha,\beta$ are learnable scale and bias parameters. For modality $m\in\{s,y\}$, let $\mathbf{h}_{m,i}$ and $\tilde{\mathbf{h}}_{m,i}$ denote the pre-projection global and localized features, respectively. Because the localized branch is sharper and less stable during early optimization, we regularize it toward the detached global branch using predictor-based self-distillation:
    \begin{equation}
        \mathcal{L}_{\mathrm{distill}} = \frac{1}{2B}\sum_{i=1}^{B}\sum_{m\in\{s,y\}} \Big[ 1-\cos\big(\psi_m(\tilde{\mathbf{h}}_{m,i}), \mathrm{sg}(\mathbf{h}_{m,i})\big) \Big],
    \end{equation}
    where $\psi_m$ is a predictor and $\mathrm{sg}(\cdot)$ is the stop-gradient operator.
    
    Relying only on exact-pair supervision may still permit a coarse absorber-dominated partition of the latent space. Let $c_i$ denote the absorber species of sample $i$, and define the off-diagonal same-absorber set as $\mathcal{P}_{\mathrm{abs}}=\{(i,j)\mid c_i=c_j,\ i\neq j\}$. Thus, we introduce an absorber-aware auxiliary regularizer on the global-stream logits:
    \begin{equation}
        \mathcal{L}_{\mathrm{absorber}} = -\frac{1}{|\mathcal{P}_{\mathrm{abs}}|} \sum_{(i,j)\in\mathcal{P}_{\mathrm{abs}}} \log \sigma \big(\alpha\,\mathbf{z}_{s,i}^{\top}\mathbf{z}_{y,j}+\beta\big).
    \end{equation}
    Unlike $\mathcal{L}_{\mathrm{global}}$, this term treats same-absorber off-diagonal pairs as auxiliary positives to soften the repulsion they receive under exact-pair supervision, thereby contracting the coarse absorber-level manifold. Importantly, $\mathcal{L}_{\mathrm{absorber}}$ does not by itself impose within-absorber margins; instead, it prevents element identity from becoming the final decision boundary, forcing the instance-matching terms $\mathcal{L}_{\mathrm{global}}$ and $\mathcal{L}_{\mathrm{fg}}$ to rely on residual coordination differences for exact retrieval. Accordingly, the desired intra-element coordination structure emerges from this joint objective rather than from any single component in isolation. The XASLip objective is
    \begin{equation}
    \mathcal{L}_{\mathrm{XASLip}} = \mathcal{L}_{\mathrm{global}} + \lambda_{\mathrm{fg}}\mathcal{L}_{\mathrm{fg}} + \lambda_{\mathrm{dist}}\mathcal{L}_{\mathrm{distill}} + \lambda_{\mathrm{abs}}\mathcal{L}_{\mathrm{absorber}}.
    \end{equation}
    Here, $\lambda_{\mathrm{fg}}$, $\lambda_{\mathrm{dist}}$, and $\lambda_{\mathrm{abs}}$ are weighting hyperparameters (ablation analyses and specific values are detailed in Section 5.4 and SI Section D). During inference and downstream retrieval, only the stable global representations $(\mathbf{z}_s,\mathbf{z}_y)$ are retained; the localized branch and shallow predictors serve purely as train-time regularizers. Overall, XASLip should be understood as an XAS-specialized alignment recipe whose gains stem from physics-aware spectral tokenization and absorber-aware optimization within a pairwise instance-matching framework, rather than from merely adopting a generic dual-stream contrastive backbone.
        
    \subsection{Anchored Absolute-Spectrum Prediction}\label{sec:forward}
    
    Uni-XAS redefines forward modeling as computing the joint target $\mathbf{Y} = [\tilde{E}_0, \mathbf{I}]$, integrated with a Retrieval-Augmented Generation mechanism to provide useful physical priors.
    
    \paragraph{Decomposed Physical Prediction.}
    To address severe numerical imbalances, we isolate the prediction of the scalar anchor from the high-variance intensity profile. Alongside a dedicated linear head regressing the normalized scalar anchor $\hat{\tilde{E}}_0$, we predict the intensity through a physically constrained decomposition. We first generate an unconstrained morphological vector $\hat{\mathbf{I}}_{\text{raw}}$ and standardize it to zero mean and unit variance to yield a scale-invariant shape $\hat{\mathbf{I}}_{\text{shape}}$. The final intensity is reconstructed using a predicted scalar gain $\hat{g}$ and a baseline bias $\hat{b}$:
    \begin{equation}
            \hat{\mathbf{I}} = \hat{g} \cdot \hat{\mathbf{I}}_{\text{shape}} + \hat{b}, \quad \text{s.t.} \quad \mathbb{E}[\hat{\mathbf{I}}_{\text{shape}}] = 0, \ \mathrm{Var}(\hat{\mathbf{I}}_{\text{shape}}) = 1.
    \end{equation}
    This explicit statistical constraint eliminates mathematical overparameterization. By separately supervising the anchor $\hat{\tilde{E}}_0$ and the decomposed intensity parameters using Mean Squared Error against their derived ground-truth counterparts, we isolate complex near-edge gradients from baseline shifts, significantly stabilizing the regression landscape.
    
    \paragraph{Retrieval-Augmented Cross-Attention Decoding.}
    We employ a Transformer-based Cross-Attention decoder to fuse physical priors into the parametric generation. Using the structural query $\mathbf{z}_s$, we perform a $K$-nearest neighbor search over the pre-computed training embeddings within the shared latent space $\mathcal{Z}$. By computing cosine similarities, we fetch the top-$K$ spectral targets. As detailed in Section~\ref{sec:experiments}, any candidate embedding sharing the identical sample index as the query is explicitly masked prior to selection, preventing trivial self-retrieval. To further mitigate the risk of degenerating into a pure retrieval system when exposed to highly similar local environments, the cross-attention outputs are injected into the main stream via a zero-initialized residual gate. This architectural bottleneck ensures the decoder initially optimizes as an independent parametric predictor, utilizing retrieved knowledge only as auxiliary guidance. Because the retrieval pool is restricted to the training split and search operates over a frozen, low-dimensional latent space, the additional retrieval overhead remains low relative to the encoder--decoder forward computation.
    
    \paragraph{Latent Manifold Consistency.}
    To enforce cross-modal semantic fidelity and intra-modal structural correctness, the predicted spectrum is re-encoded through the frozen XASLip spectral encoder to obtain $\hat{\mathbf{z}}_y$. We penalize its deviation from both the structure query $\mathbf{z}_s$ and the ground-truth embedding $\mathbf{z}_y$:
    \begin{equation}
        \mathcal{L}_{\text{align}} = \gamma_1 \big(1 - \cos(\hat{\mathbf{z}}_y, \mathbf{z}_s)\big) + \gamma_2 \big(1 - \cos(\hat{\mathbf{z}}_y, \mathbf{z}_y)\big).
    \end{equation}
    This constraint acts as a stable semantic critic without representation drift. The total forward training objective combines this alignment constraint with the aforementioned scalar anchor and decomposed intensity regression losses.
    
    \subsection{PR-Flow for Composition-Conditional Coordinate Generation}\label{sec:inverse}
    
    \paragraph{Inverse Modeling as Composition-Conditional Coordinate Generation.}
    Because mapping a 1D spectrum to a 3D local environment is inherently ill-posed, we study a controlled setting in which the absorber-centered atom types $\mathcal{V}$ and node count $N$ are provided by the benchmark protocol, and the model predicts only coordinates $\mathbf{X}$. The inverse task is therefore formulated as $p(\mathbf{X}\mid\mathbf{Y},\mathcal{V})$ rather than unconstrained full graph generation, isolating the geometric ambiguity of spectrum-conditioned local-coordinate recovery before tackling composition discovery.
    
    \paragraph{Conditioning via Latent Prototypes.}
    Given a query spectrum $\mathbf{Y}$, we encode it into $\mathbf{z}_y = \Phi_y(\mathbf{Y})$. Operating entirely within the frozen shared space $\mathcal{Z}$, we perform cross-modal retrieval to fetch the top-$K$ structural prototypes by computing the cosine similarity $\cos(\mathbf{z}_y, \mathbf{z}_s^{(i)})$ against the training structure embeddings. We enforce the same identical-sample masking strategy during training and strictly restrict the retrieval pool to the training split during inference. A cross-attention latent module aggregates the spectrum query and these retrieved prototypes into a conditioning sequence $\mathbf{Z}_{\text{cond}}$ that narrows the generative search space to structurally compatible local manifolds.
    
    \paragraph{Permutation-Rectified Flow via Type-wise Optimal Transport.}
    We instantiate the explicit 3D generator using conditional flow matching on mean-centered coordinates. Although physically cropped around the absorber, inverse training coordinates are translated to a mean-centered frame, removing global translation and focusing on relative geometry. Let $\bar{\mathbf{X}}_1 \in \mathbb{R}^{N \times 3}$ denote the centered target coordinates and let $\bar{\mathbf{X}}_0$ denote centered Gaussian noise. In local atomic environments, identical ligands are permutationally invariant. Applying flow matching over arbitrary dataset indices forces noise particles to travel along severely intersecting trajectories, triggering conflicting gradients and a severe centroid-collapse issue. To address this, we propose Permutation-Rectified Flow Matching (PR-Flow), which establishes a deterministic intra-sample coupling via type-wise optimal transport. For each element type $c$ with $N_c$ atoms, we find the optimal permutation $\pi_c^*$ that minimizes the discrete spatial 2-Wasserstein distance between the centered noise and centered target point sets:
    \begin{equation}
        \pi_c^* = \mathop{\arg\min}_{\pi \in \Pi(N_c)} \sum_{i=1}^{N_c} \left\| \bar{\mathbf{X}}_{0, i}^{(c)} - \bar{\mathbf{X}}_{1, \pi(i)}^{(c)} \right\|^2.
    \end{equation}
    Applying these optimal type-wise permutations yields the aligned target $\bar{\mathbf{X}}_1^{\text{aligned}}$. The resulting straight-line interpolation path is
    \begin{equation}
        \bar{\mathbf{X}}_t = (1-t)\bar{\mathbf{X}}_0 + t\bar{\mathbf{X}}_1^{\text{aligned}}, \qquad t \in [0,1].
    \end{equation}
    To construct the generative flow, the decoder estimates the constant target velocity $\bar{\mathbf{X}}_1^{\text{aligned}} - \bar{\mathbf{X}}_0$ by predicting a vector field $\mathbf{v}_{\theta}(\bar{\mathbf{X}}_t, t, \mathbf{Z}_{\text{cond}})$ through lightweight $E(3)$-equivariant updates. These updates are constructed by modulating the relative displacement vectors $\bar{\mathbf{X}}_{t,j} - \bar{\mathbf{X}}_{t,i}$ using invariant scalar weights derived from pairwise distances and node embeddings. This mathematical design avoids the computational overhead associated with high-order parameterizations such as spherical harmonics, while preserving geometric sensitivity. The corresponding training objective is
    \begin{equation}
        \mathcal{L}_{\text{flow}} = \mathbb{E}_{t, \bar{\mathbf{X}}_0, \bar{\mathbf{X}}_1} \left\| \mathbf{v}_{\theta}(\bar{\mathbf{X}}_t, t, \mathbf{Z}_{\text{cond}}) - (\bar{\mathbf{X}}_1^{\text{aligned}} - \bar{\mathbf{X}}_0) \right\|^2.
    \end{equation}
    Because equivariance is architecturally guaranteed by the update rule, continuous flow targets can be supervised directly in the mean-centered sample frame without explicit rotation canonicalization, inherently maintaining geometric consistency.

\section{Experiments}\label{sec:experiments}

    \subsection{Experimental Setup}
    We evaluate Uni-XAS on our curated benchmark of 328,839 structure--spectrum pairs (Section~\ref{sec:data}). We study three coupled tasks—retrieval alignment, forward prediction, and inverse generation—reporting all metrics under a strictly standardized protocol. To ensure full reproducibility, the dataset and codebase will be publicly released.
    
    \paragraph{Spectrum Protocol, Leakage Control and Implementation.}
    Spectra are represented by an absorber-normalized edge anchor $\tilde{E}_0$ and absolute intensity $\mathbf{I}\in\mathbb{R}^{128}$; anchor errors are reported after denormalizing $\tilde{E}_0$ back to the absolute physical energy scale in eV. We employ an 80/20 train/test stratified split per absorber species, drawing validation sets exclusively from training data. To prevent trivial leakage during retrieval, exact index matches are masked, and test samples are strictly excluded from the memory bank. Training follows the three-stage curriculum (Section~\ref{sec:methods}) using AdamW with mixed precision. Inverse inference uses Euler integration.

    \begin{figure*}[t]
        \centering
        \includegraphics[width=0.99\linewidth]{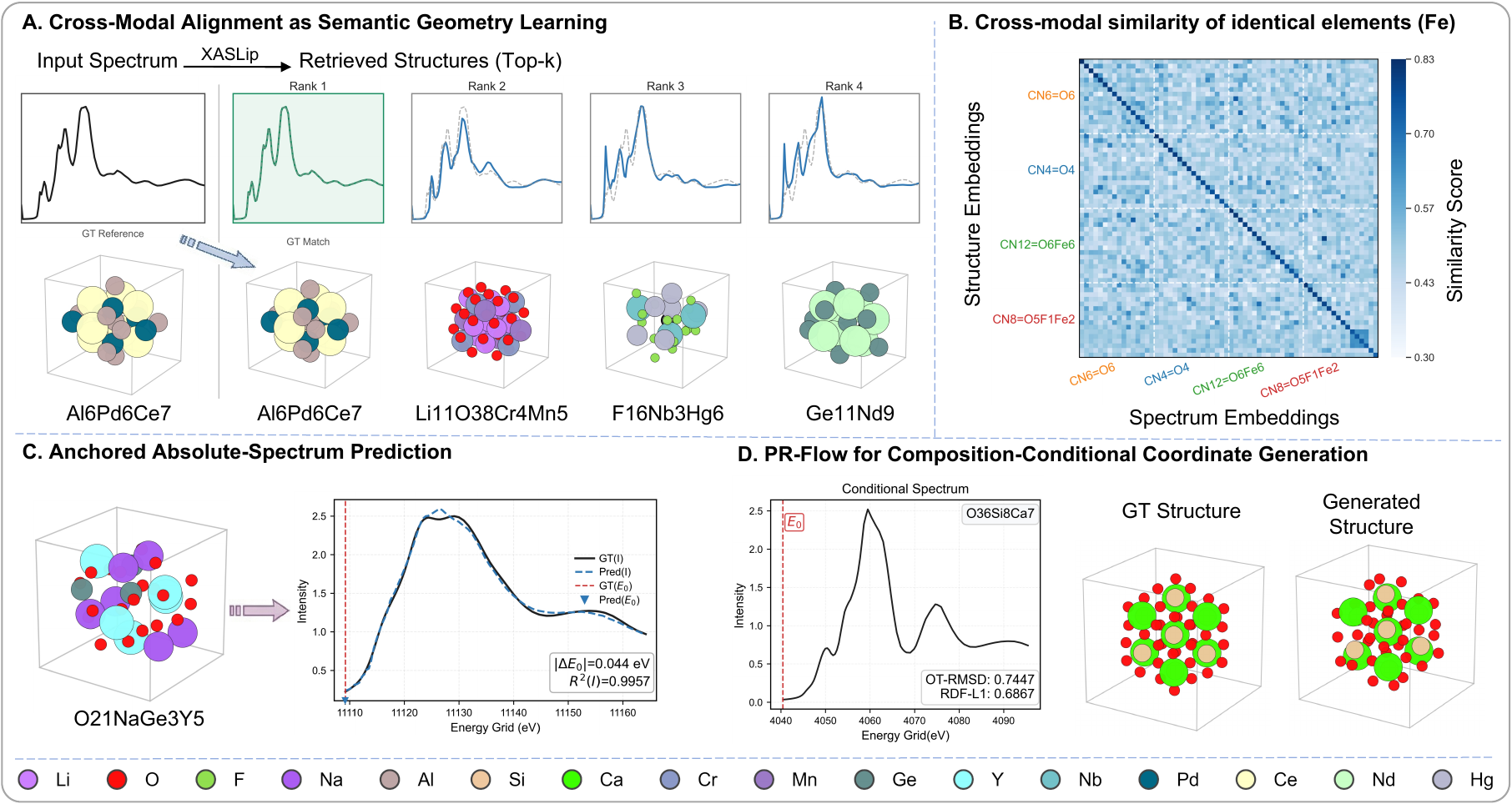}
        \caption{Qualitative results of Uni-XAS. The shared latent space effectively aligns modalities to retrieve structurally consistent 3D geometries from spectral queries (A), and resolves distinct intra-element coordination numbers within the learned manifold (B). This unified interface enables accurate forward absolute-spectrum prediction (C) and high-fidelity inverse composition-conditional generation (D). Grey boxes denote absorber-centered local crops rather than periodic unit cells.}
        \label{fig:main_results}
    \end{figure*}
    \paragraph{Baselines and Protocol Fairness.}
    We evaluate Uni-XAS against representative baselines under our standardized protocol. For retrieval, CLIP, SigLIP, and SigLIP2 objectives are instantiated with identical encoders. Forward baselines (OmniXAS~\cite{kharel2025omnixas}, XAStruct~\cite{wang2025xastruct}, CGXAS~\cite{lin2026cgxas}) are adapted to regress the joint $[\tilde{E}_0, \mathbf{I}]$ target. For inverse generation, lacking established 3D generative baselines for XAS, we introduce spectral conditioning to generative frameworks (EDM~\cite{hoogeboom2022edm}, DiffCSP-PP~\cite{jiao2024space}, IDFlow~\cite{zhou2025energy}, FlowMol3~\cite{dunn2508flowmol3}). 
    While we utilize the official implementations for these reference methods, their input and output interfaces were minimally adapted to align with our unified benchmark targets. Thus, they serve as strictly protocol-aligned controls under a shared data split policy. To ensure fair comparisons against these purely parametric baselines, we report Uni-XAS without retrieval augmentation (w/o RAG), treating non-parametric gains as an internal ablation. Full architectures and tuning constraints are detailed in the SI Section D and Section E.
    
    \subsection{Evaluation Metrics}
    
    \paragraph{Retrieval Alignment and Forward Prediction.}
    For cross-modal retrieval, we assess bidirectional Recall@K and Mean Reciprocal Rank (MRR), using average MRR for checkpoint selection. For forward prediction, we evaluate the energy coordinate and intensity morphology separately. We report MAE and MSE for the denormalized edge anchor $E_0$ in physical eV. For the intensity profile $\mathbf{I}$, we report MAE, MSE, and the Coefficient of Determination ($R^2$).
    
    \paragraph{Inverse 3D Generation.}
    Because inverse XAS mapping is inherently one-to-many, standard RMSD erroneously penalizes geometrically valid predictions due to arbitrary index mismatches among identical ligands. We therefore evaluate geometric fidelity using Optimal Transport RMSD (OT-RMSD), which jointly minimizes over type-consistent permutations $\pi \in \Pi_{\mathrm{type}}$ and rigid-body alignment $(\mathbf{R} \in \mathrm{SO}(3), \boldsymbol{\tau})$ to yield a physically meaningful geometric fidelity score~\cite{grave2019unsupervised,xu2022geodiff,liang2024foundations}. To assess macroscopic distributional realism, we compute the $L_1$ discrepancy of the Radial Distribution Functions (RDF-L1). Mathematical formulations are detailed in SI Section G.

\subsection{Main Results}
    We evaluate Uni-XAS across the three coupled tasks under our unified benchmark protocol. The alignment quality within the shared interface directly dictates the downstream performance across both forward and inverse tasks.
    
    \paragraph{Alignment in the Shared Latent Interface.}
    Table~\ref{tab:clip_retrieval_main} reports global bidirectional retrieval in the shared latent space $\mathcal{Z}$. Under this protocol, XASLip consistently improves both retrieval directions over CLIP, SigLIP, and SigLIP2 baselines, raising structure-to-spectrum R@1 from 0.3171 (SigLIP2) to 0.3800 and the average bidirectional MRR from 0.4759 to 0.5405. While macroscopic elemental separation is a fundamental prerequisite, the core challenge lies in resolving fine-grained local geometries. Panel A of Figure~\ref{fig:main_results} demonstrates this semantic alignment through accurate top-$k$ structural retrievals from spectral queries. Furthermore, Panel B visually confirms this intra-element sensitivity by illustrating the cross-modal similarity matrix exclusively for iron (Fe) samples. The clear block-diagonal structure demonstrates that the learned representation resolves fine-grained intra-element variations, successfully discriminating distinct coordination numbers (CN) within the same element. Finally, stricter within-absorber and strictly-isomeric diagnostics preserve these advantages, as detailed in Section F of the SI, confirming that XASLip effectively aligns subtle geometric perturbations rather than merely clustering macroscopic elements.
    \begin{table}[t]
        \centering
        \caption{Bidirectional cross-modal retrieval under the shared protocol. All metrics are higher the better ($\uparrow$).}
        \label{tab:clip_retrieval_main}
        
        \begin{tabular}{l cccc cccc}
            \toprule
            \multirow{2}{*}{Method} & \multicolumn{4}{c}{Structure-to-Spectrum (S2P)} & \multicolumn{4}{c}{Spectrum-to-Structure (P2S)} \\
            \cmidrule(lr){2-5} \cmidrule(lr){6-9}
            & R@1 & R@5 & R@10 & MRR & R@1 & R@5 & R@10 & MRR \\
            \midrule
            CLIP    & 0.2726 & 0.6304 & 0.7672 & 0.4324 & 0.2897 & 0.6604 & 0.7893 & 0.4534 \\
            SigLIP  & 0.3113 & 0.6661 & 0.7818 & 0.4682 & 0.3064 & 0.6699 & 0.7894 & 0.4672 \\
            SigLIP2 & 0.3171 & 0.6826 & 0.7993 & 0.4780 & 0.3108 & 0.6807 & 0.8022 & 0.4737 \\
            \midrule
            \textbf{XASLip (Ours)} & \textbf{0.3800} & \textbf{0.7435} & \textbf{0.8483} & \textbf{0.5387} & \textbf{0.3815} & \textbf{0.7490} & \textbf{0.8526} & \textbf{0.5423} \\
            \bottomrule
        \end{tabular}%
        
    \end{table}
    
    \paragraph{Forward Anchored Absolute-Spectrum Prediction.}
    Table~\ref{tab:forward_main} benchmarks forward surrogate performance. To evaluate existing methods under the same absolute-spectrum protocol, we equip each baseline with an identical $E_0$ normalization scheme and retrain them to regress the unified $[\tilde{E}_0, \mathbf{I}]$ target. Compared with these purely parametric baselines, our purely parametric counterpart, Uni-XAS (w/o RAG), already yields lower anchor errors and stronger intensity reconstruction. Activating the full retrieval-calibrated decoding (Uni-XAS Full) further provides a substantial internal gain, reducing $E_0$ Mean Squared Error to 0.0238 and pushing intensity $R^2$ to 0.9067. These results support the proposed design for decoupled physical prediction and demonstrate the additional improvement of non-parametric structural priors.

    \paragraph{Explicit Inverse 3D Structural Generation.}
    Table~\ref{tab:inverse_main} evaluates the spectrum-to-structure generation task. Adapting 3D generative models to the XAS domain reveals a substantial domain gap, especially when chemically identical ligands introduce permutation ambiguity during supervision. Operating purely parametrically, Uni-XAS (w/o RAG) achieves a Best-of-5 OT-RMSD of 1.7668, significantly outperforming these adapted baselines. Coupling this permutation-rectified foundation with retrieval conditioning (Uni-XAS Full) further optimizes the OT-RMSD to 1.7510. These findings demonstrate that resolving permutation ambiguity via PR-Flow is the primary driver of generative stability, while retrieval augmentation acts as an effective complementary geometric prior.
    
\subsection{Ablation Studies and Analyses}
    Table~\ref{tab:master_ablation} ablates Uni-XAS components across all three tasks to evaluate their contributions. Notably, because type-wise optimal transport operates as offline preprocessing, it rectifies training targets without adding parameterized complexity to the flow network.
    \begin{table}[htbp]
        \centering
        \small
        \caption{Forward anchored absolute-spectrum prediction under the unified target $\mathbf{Y}=[\tilde{E}_0,\mathbf{I}]$. Errors for the energy anchor $E_0$ are measured in actual physical electronvolts (eV).}
        \label{tab:forward_main}
        \begin{tabular}{lccccc}
            \toprule
            Method & MAE($E_0$)$\downarrow$ & MSE($E_0$)$\downarrow$ & MAE($\mathbf{I}$)$\downarrow$ & MSE($\mathbf{I}$)$\downarrow$ & $R^2$($\mathbf{I}$)$\uparrow$ \\
            \midrule
            OmniXAS~\cite{kharel2025omnixas}  & 0.8032 & 1.1805 & 0.1546 & 0.1241 & 0.8600 \\
            XAStruct~\cite{wang2025xastruct}  & 0.3981 & 0.3139 & 0.1332 & 0.1101 & 0.8813 \\
            CGXAS~\cite{lin2026cgxas}     & 0.5032 & 0.4637 & 0.1267 & 0.1025 & 0.8899 \\
            \textbf{Uni-XAS (Ours, w/o RAG)} & 0.1295 & 0.0489 & 0.1200 & 0.1083 & 0.8995 \\
            \midrule
            \textbf{Uni-XAS (Ours, Full)} & \textbf{0.0896} & \textbf{0.0238} & \textbf{0.1139} & \textbf{0.1006} & \textbf{0.9067} \\
            \bottomrule
        \end{tabular}
        
    \end{table}
    
\begin{table}[t]
        \centering
        \small
        \caption{Inverse local-structure generation results. Metrics are computed with rigid-motion and same-type permutation invariance. $\dagger$ denotes external generative frameworks adapted to this setting and retrained from scratch.}
        \label{tab:inverse_main}
        \begin{tabular}{lcccc}
            \toprule
            \multirow{2}{*}{Method} & \multicolumn{2}{c}{Geometry Fidelity} & \multicolumn{2}{c}{Distance Distribution} \\
            \cmidrule(lr){2-3} \cmidrule(lr){4-5}
            & OT-RMSD$_{\mathrm{Exp}}\downarrow$ & Best-of-5 OT-RMSD$\downarrow$ & RDF-L1$_{\mathrm{Exp}}\downarrow$ & Best-of-5 RDF-L1$\downarrow$ \\
            \midrule
            EDM$^{\dagger}$~\cite{hoogeboom2022edm} & 2.5405 & 2.3023 & 0.6051 & 0.5532 \\
            DiffCSP-PP$^{\dagger}$~\cite{jiao2024space} & 3.4815 & 2.6976 & 0.5343 & 0.4325 \\
            FlowMol3$^{\dagger}$~\cite{dunn2508flowmol3} & 2.4403 & 2.2424 & 0.5279 & 0.4609 \\
            IDFlow$^{\dagger}$~\cite{zhou2025energy} & 2.2323 & 2.0498 & 0.6817 & 0.6254 \\
            \textbf{Uni-XAS (Ours, w/o RAG)} & 2.0306 & 1.7668 & 0.4662 & 0.4078 \\
            \midrule
            \textbf{Uni-XAS (Ours, Full)} & \textbf{2.0222} & \textbf{1.7510} & \textbf{0.4569} & \textbf{0.3968} \\
            \bottomrule
        \end{tabular}
        
\end{table}

\begin{table}[t]
    \centering
    \small
    \caption{Ablation summary across retrieval, forward prediction, and inverse generation. Each part reports progressive variants to isolate the contribution of specific design choices.}
    \label{tab:master_ablation}
    \resizebox{\columnwidth}{!}{%
    \begin{tabular}{@{}lccccc@{}}
        \toprule
        \multicolumn{6}{c}{\textbf{Part I: Cross-Modal Alignment (Retrieval)}} \\
        \midrule
        \multicolumn{2}{@{}l}{Configuration} & S2P R@1$\uparrow$ & P2S R@1$\uparrow$ & S2P MRR$\uparrow$ & P2S MRR$\uparrow$ \\
        \midrule
        \multicolumn{2}{@{}l}{Dual-Stream Baseline} & 0.3161 & 0.3119 & 0.4782 & 0.4750 \\
        \multicolumn{2}{@{}l}{+ Physics-Aware Encoding} & 0.3484 & 0.3482 & 0.5080 & 0.5092 \\
        \multicolumn{2}{@{}l}{+ Absorber-Aware Constraint (Full XASLip)} & \textbf{0.3800} & \textbf{0.3815} & \textbf{0.5387} & \textbf{0.5423} \\
        \midrule
        \multicolumn{6}{c}{\textbf{Part II: Forward Absolute-Spectrum Prediction}} \\
        \midrule
        \multicolumn{2}{@{}l}{Configuration} & MAE($E_0$)$\downarrow$ & \multicolumn{2}{c}{MSE(I)$\downarrow$} & $R^2$(I)$\uparrow$ \\
        \midrule
        \multicolumn{2}{@{}l}{Task-Specific Regression (w/o Retrieval)} & 0.1338 & \multicolumn{2}{c}{0.1296} & 0.8848 \\
        \multicolumn{2}{@{}l}{+ Decomposed Physical Prediction} & 0.1295 & \multicolumn{2}{c}{0.1083} & 0.8995 \\
        \multicolumn{2}{@{}l}{+ Retrieval-Augmented Cross-Attention} & 0.2647 & \multicolumn{2}{c}{0.1031} & 0.9045 \\
        \multicolumn{2}{@{}l}{+ Latent Manifold Consistency (Full Uni-XAS)} & \textbf{0.0896} & \multicolumn{2}{c}{\textbf{0.1006}} & \textbf{0.9067} \\
        \midrule
        \multicolumn{6}{c}{\textbf{Part III: PR-Flow for Explicit Inverse Structure Generation}} \\
        \midrule
        Model Variant & RAG Conditioning & \multicolumn{2}{c}{Permutation Rectification} & Best-of-5 OT-RMSD$\downarrow$ & Best-of-5 RDF-L1$\downarrow$ \\
        \midrule
        Centered Flow (Eq. Updates)  & $\times$ & \multicolumn{2}{c}{$\times$} & 1.8903 & 0.4493 \\
        Centered Flow (Eq. Updates) + RAG  & $\checkmark$ & \multicolumn{2}{c}{$\times$} & 3.2368 & 1.7657 \\
        Centered Flow (Eq. Updates) + PR   & $\times$ & \multicolumn{2}{c}{$\checkmark$} & 1.7668 & 0.4078 \\
        \textbf{Uni-XAS (Full)} & $\checkmark$ & \multicolumn{2}{c}{$\checkmark$} & \textbf{1.7510} & \textbf{0.3968} \\
        \bottomrule
    \end{tabular}
    }
\end{table}
    
    \paragraph{Dissecting XASLip Alignment.}
    Part I of Table~\ref{tab:master_ablation} ablates the cross-modal representation backbones, demonstrating that upgrading the generic baseline to a Physics-Aware Encoding improves the S2P R@1 from 0.3161 to 0.3484, supporting the utility of explicitly capturing onset sharpness and peak curvature. Integrating the Absorber-Aware Constraint further increases the S2P and P2S MRRs to 0.5387 and 0.5423, respectively. As corroborated by our intra-element diagnostics (SI Section F), this constraint is crucial for distinguishing subtle coordination geometries, ensuring the learned space reflects fine-grained structural variations rather than defaulting to macroscopic elemental clustering.
    
    \paragraph{Deconstructing Forward Prediction.}
    Part II evaluates forward prediction, starting with an isolated parametric baseline. Introducing Decomposed Physical Prediction factorizes the intensity scale, reducing MSE from 0.1296 to 0.1083 while preserving a stable energy anchor. Integrating Retrieval-Augmented Cross-Attention leverages spectral priors to refine the intensity profile ($R^2$ reaches 0.9045), but makes the absolute energy anchor vulnerable to drift ($E_0$ MAE increases to 0.2647). Adding Latent Manifold Consistency addresses this: regularizing predictions in the frozen XASLip space corrects the drift, dropping MAE to 0.0896 while maintaining improved intensity reconstruction.
    
    \paragraph{Interaction Between Retrieval Conditioning and Permutation Rectification.}
    Part III highlights a non-trivial interaction during explicit inverse generation. We establish a solid foundation using a centered flow decoder with $E(3)$-equivariant coordinate updates (Eq. Updates), a purely parametric spectrum-conditioned flow-matching architecture specifically designed for this task (SI Section D). This custom base model alone yields an OT-RMSD of 1.8903. Notably, directly incorporating retrieval-augmented latent conditioning leads to a severe performance drop, increasing the OT-RMSD to 3.2368. We attribute this to ligand permutation ambiguity: when target coordinates are arbitrarily ordered, injecting retrieved structural prototypes introduces conflicting gradient signals rather than constructive geometric priors. Applying Permutation Rectification addresses this ambiguity, improving the OT-RMSD to 1.7668 compared to the base model. More importantly, once PR establishes permutation-consistent training targets, the RAG conditioning becomes consistently beneficial. The combined approach achieves the lowest Best-of-5 OT-RMSD of 1.7510 and an RDF-L1 of 0.3968. This indicates that mathematically resolving permutation ambiguity is a necessary prerequisite for effectively utilizing retrieval-augmented generation. Furthermore, consistent gains from RAG highlight its robust role as a non-parametric refinement. Within this framework, retrieved priors actively refine spectral profiles in forward prediction and provide fine-grained geometric cues for inverse generation, complementing the PR-Flow driven trajectories.
    
\subsection{Limitations and Future Work}
While Uni-XAS establishes a baseline, several limitations motivate future work. First, the benchmark and all three tasks are defined under a standardized local-graph protocol; though necessary for controlled comparison, transferring the interface to noisier experimental spectra and less curated structures remains an open sim-to-real challenge. Second, the frozen latent space and retrieval-augmented design prioritize stability, restricting task-specific adaptation and extrapolation to novel out-of-distribution environments. Third, inverse generation currently assumes known atomic compositions, leaving unconstrained structure discovery unexplored. Finally, while our centered $E(3)$-equivariant updates are computationally efficient, the generated structures lack strict rotation-canonical supervision and thermodynamic constraints. Addressing these by combining experimental adaptation, composition discovery, and physics-guided priors remains a critical direction.

\section{Conclusion}
We presented Uni-XAS, a unified bidirectional framework that establishes structure--spectrum alignment as a shared latent interface rather than a standalone pretraining objective. This cross-modal space supports three coupled capabilities: fine-grained retrieval, anchored absolute-spectrum prediction, and composition-conditional local-coordinate generation via PR-Flow. Across these tasks, our results demonstrate that reusing one aligned latent space provides a consistent foundation for multimodal representation learning in scientific spectroscopy. More broadly, we hope this work positions XAS as a meaningful testbed for advancing cross-modal alignment, retrieval-augmented prediction, and flow-based generative modeling in heterogeneous scientific data.

\section{Acknowledgment}
F.T. acknowledges the National Key R\&D Program of China (Grant No. 2025YFA1511003), National Natural Science Foundation of China (Grant No. 22573085) and a startup fund at Xiamen University. J.C. acknowledges the National Natural Science Foundation of China (Grant Nos. 22225302, 92470201, 92461312, 22541204, 22021001) and the Fundamental Research Funds for the Central Universities 20720250005, Fundamental and Interdisciplinary Disciplines Breakthrough Plan of the Ministry of Education of China JYB2025XDXM307, Laboratory of AI for Electrochemistry (AI4EC), IKKEM (Grant Nos. RD2023100101 and RD2022070501). This work used the computational resources in the IKKEM intelligent computing center.

\bibliography{main}

\appendix

\section*{\hspace{-4mm} \centering Appendix}
\vspace{3mm}

This supplementary information documents the dataset construction, split rule, model settings, baseline adaptations, and evaluation details used in the main paper. Its purpose is to make the reported results easier to inspect, especially with respect to sample pairing, checkpoint selection, retrieval-bank isolation, and metric computation. We do not introduce additional claims here; instead, we record the implementation and evaluation choices needed to read the reported numbers correctly. Sections A--C summarize the benchmark and evaluation rules, Sections D--F document model details, baseline adaptation scope, and additional diagnostics, Section G provides the mathematical definitions omitted from the main text for space, and Section H presents extended qualitative visualizations for both bidirectional tasks.

% ================================================================
% A  Reproducibility Overview
% ================================================================
\section{Reproducibility Overview}
\label{app:repro}

    The main paper studies Uni-XAS as a unified framework for three coupled tasks: cross-modal retrieval, forward anchored absolute-spectrum prediction, and inverse local-structure generation. Because the benchmark and several adapted baselines rely on the same split, target definition, and evaluator, we summarize below the task-level inputs, outputs, validation criteria, and leakage controls used throughout the study.

    \paragraph{Design principle.}
    We distinguish three kinds of choices in the appendix. First, some changes are required simply to run a baseline on our data format. Second, some settings are shared by all methods, such as the split, target definition, evaluator, and leakage-control rules. Third, hyperparameter search is kept small and fixed in advance. In direct correspondence with Section~5.1--5.2 of the main paper, retrieval checkpoints are selected by validation average bidirectional MRR, forward baselines by validation loss under the shared target $\mathbf{Y}=[\tilde{E}_0,\mathbf{I}]$, Uni-XAS forward by validation intensity $R^2$ with $E_0$ error archived jointly, and inverse models by validation OT-RMSD$_{\mathrm{BoS}}$ followed by one full-sampling re-check after the configured epoch threshold. Across CLIP, SigLIP, SigLIP2, and XASLip, encoder backbones, projection dimensionality, batch construction, and training budget are matched; only the alignment objective differs.

\begin{table*}[t]
    \centering
    \small
\caption{Reproducibility overview of the benchmark and evaluation setup. All tasks use the same absorber-stratified training / held-out evaluation split, while checkpoint selection relies on training-side monitoring rather than on the released \texttt{valid\_lmdb} split.}
    \label{tab:repro_overview}
    \resizebox{\textwidth}{!}{%
    \begin{tabular}{L{2.0cm} L{3.2cm} L{3.4cm} L{2.8cm} L{3.2cm} L{4.6cm}}
        \toprule
        Task & Input $\rightarrow$ Output & Compared Methods & Validation / Checkpoint Selection & Reported Held-out Metrics & Key Leakage / Fairness Controls \\
        \midrule
        Cross-modal retrieval &
        absorber-centered local graph $\leftrightarrow$ standardized spectrum &
        CLIP, SigLIP, SigLIP2, XASLip (all re-instantiated with the same structure and spectral encoders) &
        average bidirectional MRR, i.e., $\tfrac{1}{2}(\mathrm{MRR}^{\mathrm{S2P}}+\mathrm{MRR}^{\mathrm{P2S}})$ &
        bidirectional Recall@1/5/10 and MRR; additional within-absorber and same-node-count diagnostics &
        identical paired training / held-out evaluation split; same encoders and projection dimensionality for all retrieval objectives; no held-out evaluation data used during model selection \\
        \midrule
        Forward prediction &
        absorber-centered local graph $\rightarrow$ anchored absolute spectrum $\mathbf{Y}=[\tilde{E}_0,\mathbf{I}]$ (reported as $E_0$ after denormalization) &
        OmniXAS, XAStruct, CGXAS, Uni-XAS forward module &
        the same training-side monitoring rule; for Uni-XAS forward, validation intensity $R^2$ is used as the main score while loss and $E_0$ error are archived jointly &
        MAE/MSE on denormalized $E_0$ in eV; MAE/MSE/$R^2$ on non-normalized absolute intensity &
        identical target reformulation for all methods; same element-wise $E_0$ normalization / denormalization table; no held-out evaluation spectra in retrieval memory; identity masking during training for retrieval-enabled models \\
        \midrule
        Inverse generation &
        standardized spectrum + known local composition count $\rightarrow$ 3D coordinates &
        EDM, DiffCSP-PP, FlowMol3, IDFlow, Uni-XAS inverse generator &
        training-side monitoring; for Uni-XAS inverse, candidate checkpoints are selected by validation OT-RMSD$_{\mathrm{BoS}}$ and re-checked with full sampling after the configured epoch threshold &
        OT-RMSD$_{\mathrm{Exp}}$, OT-RMSD$_{\mathrm{BoS}}$, RDF-L1$_{\mathrm{Exp}}$, RDF-L1$_{\mathrm{BoS}}$ with $S=5$ hypotheses &
        all external inverse baselines remain purely parametric and do not access the Uni-XAS retrieval memory bank; unified evaluator, same number of hypotheses, same type-consistent permutation handling, and training-bank-only retrieval for Uni-XAS \\
        \bottomrule
    \end{tabular}
    }
\end{table*}

% ================================================================
% B  Dataset and Split Protocol
% ================================================================
\section{Dataset Construction, Filtering, and Split Protocol}
\label{app:data}

\subsection{Cross-Modal Pairing and Absorber-Centric Local Graphs}
    \label{app:data_pairing}

    The development of multimodal foundation models requires large-scale paired observations that are physically aligned at the sample level. Public materials repositories provide X-ray absorption spectra and crystal structures, but they do not provide them as paired samples with a unified split, target definition, and evaluator for bidirectional learning. This mismatch is especially severe for XAS, where the structural branch is geometric and local, whereas the spectral branch is sequential and continuous.
    
    We therefore construct a dedicated data engine that mines candidate K-edge XANES entries from the public XASDataLibrary~\cite{xasdatalibrary} and links each spectrum to its corresponding crystallographic identifier in the Materials Project~\cite{jain2013commentary} using pymatgen-based metadata reconciliation. Entries with ambiguous structure linkage, incomplete metadata, or invalid absorber specifications are filtered prior to pair formation. Following the establishment of valid structure-spectrum pairs, we perform systematic quality control on the dataset to ensure its reliability, physical consistency, and statistical representativeness: we group the samples by the elemental species of the absorbing atom, exclude element groups with fewer than 1000 samples that lack sufficient statistical representativeness, and apply the 3× interquartile range (IQR) criterion to the remaining statistically valid groups to remove outlier samples with abnormal absorption edge onset energy. We further conduct physical validity filtering on the spectral data, discarding non-physical spectra with negative absorption intensity and invalid samples without effective characteristic absorption peaks.
    
    Because XANES is primarily governed by the local scattering environment rather than the infinite periodic crystal, we transform each structure into an absorber-centered local graph using a radial cutoff of $5.0$\,\AA. This transformation shifts the problem from global crystal modeling to the local coordination regime relevant to the measured spectral signal. For inverse generation, these absorber-centric local coordinates are subsequently translated to a mean-centered frame. Thus, absorber-centered cropping defines the physically relevant neighborhood, while mean-centering provides a translation-free representation for generative learning.

    \begin{figure}[htbp]
        \centering
        \includegraphics[width=0.9\linewidth]{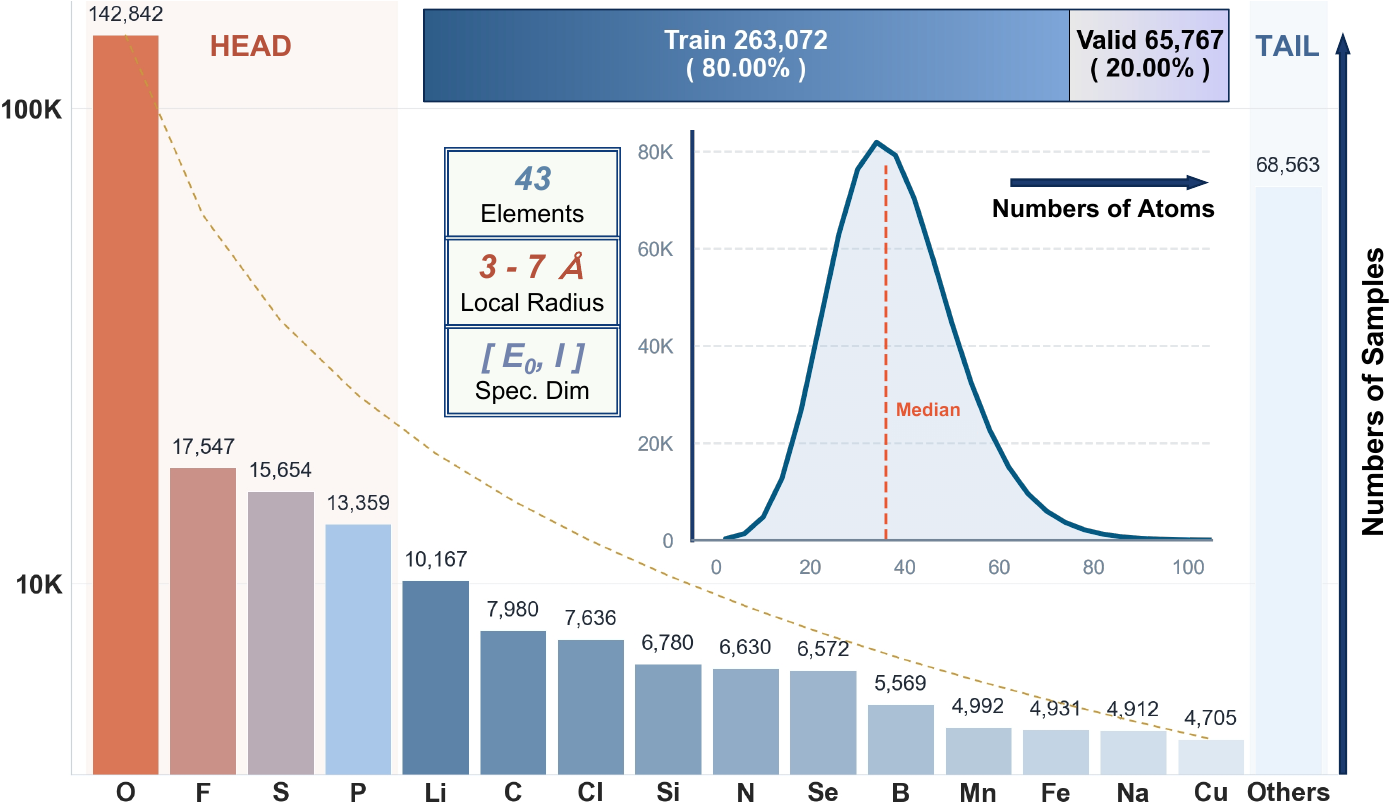}
        \caption{Statistics of the curated paired benchmark, detailing overall dataset scale, absorber distribution, and local-graph-size variability.}
        \label{fig:benchmark_overview}
    \end{figure}

    \subsection{Unified Spectral Standardization}
    \label{app:data_spec}
    Raw spectra from public repositories differ in energy range, sampling density, and calibration quality. To reduce these differences, we preprocess all spectra with the same standardization procedure before training and evaluation.
    
    For each accepted structure--spectrum pair, we first identify the physical edge anchor $E_0$ using a deterministic first-derivative criterion, which locates the point of maximal absorption rise after smoothing. We then retain the corresponding absolute intensity profile $\mathbf{I}\in\mathbb{R}^{128}$ without any global rescaling. Importantly, we deliberately avoid normalizing $\mathbf{I}$ into $[0,1]$, as such operations would remove physically meaningful amplitude information and prevent consistent absolute-intensity calibration across samples.
    
    For model training, the edge anchor is mapped into an absorber-specific normalized coordinate $\tilde{E}_0\in[0,1]$ using element-wise bounds, yielding the standardized target representation
    \[
    \mathbf{Y} = [\tilde{E}_0,\mathbf{I}].
    \]
    This results in a unified 129-dimensional target shared by all tasks and baselines during optimization.
    
    This design enforces a clear separation between scale-invariant learning (through normalized $\tilde{E}_0$) and physically grounded intensity modeling (through absolute $\mathbf{I}$). It also yields a more stringent and physically meaningful benchmark for cross-modal retrieval and inverse generation, as models must simultaneously capture spectral shape and absolute magnitude. During evaluation, all predicted anchors are rigorously converted back to the physical electronvolt (eV) domain using the same normalization table, ensuring that reported metrics remain directly interpretable in real-world units.

    \subsection{Benchmark Scale and Split Policy}
    \label{app:data_split}
    
    After filtering and standardization, the resulting benchmark contains 328{,}839 curated structure--spectrum pairs spanning 43 absorber species. To preserve elemental balance while reflecting the intrinsic long-tailed distribution of real-world materials data, we adopt an absorber-stratified split: for each absorber species, 80\% of the pairs are used for training and 20\% are reserved as a held-out evaluation split. This results in 263{,}072 training pairs and 65{,}767 held-out evaluation pairs. In the code and configuration files, this held-out LMDB is exposed under the legacy name \texttt{valid\_lmdb}; this is a naming convention inherited from the original training code rather than a different sample universe.
    
    To prevent information leakage and ensure fair generalization assessment, we perform split assignment once at the paired-sample level and keep it fixed for retrieval, forward prediction, and inverse generation. Each structure--spectrum pair is assigned to exactly one partition, and pairing relationships are preserved throughout the split. To reduce structure-level leakage, exact and symmetry-equivalent duplicate structures are removed before splitting using one fixed symmetry-aware matching procedure shared by all tasks. During training, checkpoint selection relies on training-side monitoring batches sampled from the training portion rather than on the held-out evaluation LMDB. After model selection, the chosen checkpoint is evaluated once on the held-out evaluation split (\texttt{valid\_lmdb} in the released configs). Consequently, Table~1--Table~4 in the main paper and all supplementary diagnostics are evaluated on one identical frozen held-out split rather than on task-specific sample universes. Despite these precautions, subtle structural similarities across samples originating from large public repositories cannot be completely ruled out. Such residual correlations are an inherent limitation of large-scale benchmark curation and should be considered when interpreting in-distribution generalization performance.
    
    In addition, strict isolation is enforced for retrieval-augmented components: the retrieval memory bank is constructed using training data only, and no held-out evaluation samples are accessible during either training or inference. Furthermore, no held-out evaluation metric is used at any stage of hyperparameter tuning or checkpoint selection.
    
    This benchmark uses the same input/output definition across the three tasks studied in this work---cross-modal retrieval, anchored absolute-spectrum prediction, and inverse local-structure generation---while retaining the main challenges of multimodal XAS modeling, including long-tailed absorber distributions, heterogeneous local graph sizes, and the ill-posed one-to-many nature of spectral inversion.

% ================================================================
% C  Unified Protocol
% ================================================================
\section{Unified Training and Evaluation Protocol with Leakage Control}
\label{app:protocol}

    \subsection{Common Runtime Environment}
    \label{app:runtime}
    
    Unless otherwise stated, all experiments are conducted using automatic mixed precision (AMP) in float16. Optimization follows AdamW for Uni-XAS, while adapted baselines use their source-consistent optimizers with only limited tuning. The primary results reported in Section~5 of the main paper are single-seed results under seed 42. To separate checkpoint selection from final reporting, training uses internal monitoring batches drawn from the training portion, and each selected configuration is then evaluated once on the held-out evaluation split exposed as \texttt{valid\_lmdb} in the code package. Additional repeated runs were used only as sanity checks of ranking stability and were not used to select or replace the reported checkpoints.
    
    \subsection{Shared Target Representation and Physical Evaluation}
    \label{app:target_protocol}
    All forward models operate on a unified standardized target:
    \[
    \mathbf{Y} = [\tilde{E}_0,\mathbf{I}],
    \]
    where $\tilde{E}_0\in[0,1]$ denotes the absorber-specific normalized edge anchor and $\mathbf{I}\in\mathbb{R}^{128}$ is the absolute intensity profile.
    
    A shared lookup table of absorber-specific energy ranges (provided as \texttt{Energy\_Ele\_Norm.csv}) is used consistently across all methods to normalize $E_0$ during training and to convert predictions back to the physical electronvolt (eV) domain during evaluation.
    
    All reported forward metrics are computed after denormalization:
    \begin{itemize}[leftmargin=*,nosep]
        \item MAE and MSE on $E_0$ in eV,
        \item MAE, MSE, and $R^2$ on the non-normalized intensity $\mathbf{I}$.
    \end{itemize}
    
    For inverse generation, all methods are evaluated using a shared implementation of the evaluation pipeline with a fixed number of hypotheses $S=5$ per held-out evaluation sample. Geometry fidelity and distributional realism are measured using OT-RMSD and RDF-L1, respectively (see Sec.~\ref{app:math_metrics}). All evaluation metrics are computed using a unified implementation to eliminate discrepancies arising from independent metric definitions.
    
    Model selection uses training-side monitoring with the same predefined validation rule, and no held-out evaluation metric is used for tuning or early stopping. All reported results are obtained from a single final evaluation pass on the held-out evaluation split.
    
    \subsection{Leakage Prevention and Retrieval-Bank Isolation}
    \label{app:leakage}
    
    Because Uni-XAS incorporates retrieval-augmented conditioning, leakage control follows three rules consistent with Section~5.1 of the main paper: (i) exact sample-ID matches are masked at training time, (ii) the retrieval bank is built from training data only and excludes all held-out evaluation samples, and (iii) the underlying split follows the structure-deduplicated procedure described in Section~B. As a result, the retrieval-augmented results in Table~2 and Table~3 use train-bank retrieval under split isolation rather than any transductive access to held-out data.
    
    \paragraph{Training-time identity masking.}
    Each paired sample is assigned a unique identifier stored alongside its embedding in the retrieval memory bank. During retrieval, any entry sharing the same identifier as the query is masked by assigning its similarity score to $-\infty$ before top-$K$ selection. This prevents a query from retrieving its own paired target and avoids trivial memorization through retrieval. Retrieved information is injected via a zero-initialized residual gate, ensuring that the model begins as a purely parametric predictor and gradually learns to incorporate non-parametric signals.
    
    \paragraph{Held-out evaluation-time retrieval isolation.}
    During held-out evaluation, the retrieval memory bank is constructed strictly from the training data. No embeddings from the held-out evaluation split are included, and no held-out evaluation information is accessible during retrieval or conditioning. This keeps retrieval train-only and prevents cross-split information flow.
    
    \paragraph{Fair comparison with baselines.}
    All external inverse baselines are kept purely parametric and do not access the retrieval memory bank. The principal parametric comparison is against Uni-XAS (w/o RAG), while the gap from Uni-XAS (w/o RAG) to Uni-XAS (Full) reflects the additional effect of retrieval augmentation within the same backbone family.

% ================================================================
% D  Method Details and Hyperparameters
% ================================================================
\section{Additional Method Details and Hyperparameters}
\label{app:impl}

    \subsection{Structure Encoder Architecture}\label{app:structure_encoder}
    Within the Uni-XAS framework, the structure encoder $\Phi_s$ serves as the geometric interface for processing the absorber-centered local graph $\mathbf{G}=(\mathbf{V},\mathbf{X})$. To obtain robust representations of local 3D coordination environments, we instantiate $\Phi_s$ using an $SE(3)$-aware geometric Transformer backbone based on Uni-Mol~\cite{zhou2023uni}, initialized with publicly available pre-trained weights for inorganic crystals. This design provides strong geometric priors for the subsequent cross-modal alignment and generative stages.
    
    The backbone comprises $L_{\mathrm{struct}}=15$ Transformer layers with hidden dimension $d_{\mathrm{model}}=512$, feed-forward dimension $d_{\mathrm{ff}}=2048$, and $64$ attention heads. To encode local geometry induced by the coordinates $\mathbf{X}\in\mathbb{R}^{N\times 3}$, pairwise inter-atomic distances are expanded using $64$ Gaussian radial basis functions up to a maximum cutoff of $5.0$\,\AA. These distance features are injected into self-attention as geometry-aware bias terms.
    
    The input sequence is organized as $[\mathrm{CLS}, \mathrm{atom}_1, \dots, \mathrm{atom}_N]$, where the absorber is placed at the first atomic position by construction. The $\mathrm{CLS}$ token serves as a learnable global aggregator. Its final hidden state, denoted as $\mathbf{g}_s \in \mathbb{R}^{d_{\mathrm{model}}}$, is used as the global pre-projection structural feature, consistent with the notation in the main Methods. In parallel, the encoder outputs the token-level hidden states
    \begin{equation}
    \mathbf{T}_s = \{\mathbf{h}_n\}_{n=1}^N,
    \end{equation}
    which are used by the fine-grained alignment branch via Multi-Head Attention Pooling (MAP). In practice, $\Phi_s$ produces the global feature $\mathbf{g}_s$ and token-level features $\mathbf{T}_s$, from which the shared latent embeddings $\mathbf{z}_s$ and $\tilde{\mathbf{z}}_s$ are constructed as defined in Sec.~4.2.
    
    To further emphasize the immediate coordination environment, we define an auxiliary absorber-aware feature by Gaussian distance-weighted pooling over the atomic token representations:
    \begin{equation}
        \mathbf{h}^{\mathrm{aux}}_{s}
        =
        \frac{\sum_{n=1}^{N} w_n \mathbf{h}_n}{\sum_{n=1}^{N} w_n},
        \qquad
        w_n=\exp\!\left(-\frac{d_{n,\mathrm{abs}}^2}{2\sigma_s^2}\right),
    \end{equation}
    where $\mathbf{h}_n$ is the final hidden state of atom $n$, $d_{n,\mathrm{abs}}$ denotes its Euclidean distance to the absorber, and $\sigma_s$ is the locality scale, set to $2.5$\,\AA\ in the main configuration. The $\mathrm{CLS}$ token is excluded from this pooling operation.
    
    This auxiliary feature $\mathbf{h}^{\mathrm{aux}}_{s}$ provides a soft locality-biased summary that complements the global representation $\mathbf{g}_s$. In the XASLip alignment stage, only $\mathbf{g}_s$ and $\mathbf{T}_s$ are used to define the dual-stream matching branches, whereas $\mathbf{h}^{\mathrm{aux}}_{s}$ is retained solely as an auxiliary locality-aware signal for diagnostics and ablation analysis.
    
    \subsection{Spectral Encoder Architecture}\label{app:spectral_encoder}
    The spectral encoder $\Phi_y$ provides the implementation details corresponding to the physics-aware spectral representation described in Sec.~4.2. It is designed to capture both global spectral characteristics and fine-grained near-edge structures through a multi-scale formulation.
    
    \paragraph{Physics-aware tokenization.}
    Following the design in the main text, we construct a multi-channel spectral sequence by augmenting the absolute intensity $\mathbf{I}\in\mathbb{R}^{128}$ with its first- and second-order finite differences computed after Gaussian smoothing. Specifically, let $\mathbf{I}^{(s)} = \mathbf{I} * G_{\sigma}$ denote the smoothed signal. The derivatives are defined as:
    \begin{equation}
        d^{(1)}_t = \frac{I^{(s)}_{t+1}-I^{(s)}_{t-1}}{2}, \qquad
        d^{(2)}_t = I^{(s)}_{t+1}-2I^{(s)}_{t}+I^{(s)}_{t-1}.
    \end{equation}
    The resulting three-channel sequence $[\mathbf{I}, \mathbf{d}^{(1)}, \mathbf{d}^{(2)}]$ is concatenated and prepended with a dedicated anchor token encoding $\tilde{E}_0$, yielding an input sequence of length $L=129$. This sequence is linearly projected to dimension $d_{\mathrm{model}}=128$ to form $\mathbf{X}^{(0)} \in \mathbb{R}^{L\times d_{\mathrm{model}}}$.
    
    \paragraph{Convolution-attention backbone.}
    The sequence is processed by $K=3$ stacked blocks combining local convolution and global self-attention, as described in the main text. Each layer is defined as:
    \begin{align}
        \tilde{\mathbf{X}}^{(l)} &= \mathbf{X}^{(l-1)} + \mathrm{PW}\!\left(\mathrm{SiLU}\!\left(\mathrm{DW}\!\left(\mathrm{LN}\!\left(\mathbf{X}^{(l-1)}\right)\right)\right)\right), \\
        \mathbf{X}^{(l)} &= \tilde{\mathbf{X}}^{(l)} + \mathrm{FFN}\!\left(\mathrm{LN}\!\left(\tilde{\mathbf{X}}^{(l)} + \mathrm{MHSA}\!\left(\mathrm{LN}\!\left(\tilde{\mathbf{X}}^{(l)}\right)\right)\right)\right),
    \end{align}
    where $\mathrm{DW}$ and $\mathrm{PW}$ denote depth-wise and point-wise 1D convolutions, respectively. This hybrid design captures both local spectral morphology and long-range dependencies.
    
    \paragraph{Edge-onset localization and multi-scale outputs.}
    We define the edge-onset index as:
    \begin{equation}
        t^{\star}=\arg\max_t d^{(1)}_t,
    \end{equation}
    which identifies the steepest absorption rise.
    
    Given the final hidden states $\mathbf{X}^{(K)} = [\mathbf{x}_{\mathrm{anchor}}^{(K)}, \mathbf{x}_1, \dots, \mathbf{x}_{128}]$, the encoder produces three outputs aligned with the dual-stream alignment design:
    
    \begin{itemize}[leftmargin=*,nosep]
        \item \textbf{Global embedding:} $\mathbf{z}_y = \mathrm{MeanPool}(\mathbf{X}^{(K)})$, corresponding to the global spectral representation used for stable cross-modal retrieval.
        
        \item \textbf{Token-level features:} the full sequence $\mathbf{X}^{(K)}$, which is fed into the MAP-based fine-grained alignment branch.
        
        \item \textbf{Auxiliary edge-window summary:}
        \begin{equation}
            \mathbf{z}^{\mathrm{edge}}_y = \mathrm{MeanPool}\big(\{\mathbf{x}_t \mid |t-t^{\star}|\le w\}\big),
        \end{equation}
        which provides an auxiliary edge-onset-guided local summary. In the main configuration, the window size is 7. In the implemented XASLip objective, the fine-grained branch is obtained separately by applying MAP to the returned token-level features rather than by directly using $\mathbf{z}^{\mathrm{edge}}_y$ as the localized contrastive embedding.
    \end{itemize}
    
    The auxiliary edge-window summary $\mathbf{z}^{\mathrm{edge}}_y$ provides a coordination-sensitive near-edge signal, while $\mathbf{z}_y$ captures the global absorption profile. In the implemented XASLip dual-stream alignment, the global feature and token-level features define the two contrastive branches, whereas $\mathbf{z}^{\mathrm{edge}}_y$ is retained as an auxiliary physically grounded summary.
    
    \subsection{XASLip Alignment Stage}
    \label{app:xaslip_train}
    
    This subsection provides the implementation details of the XASLip alignment stage introduced in Sec.~4.3. Consistent with the main text, the goal of this stage is to construct a shared cross-modal latent space by combining global instance-level matching, fine-grained localized matching, predictor-based self-distillation, and absorber-aware manifold regularization.
    
    XASLip is trained for 500 epochs with batch size 128, learning rate $10^{-3}$, weight decay $10^{-4}$, and gradient clipping at norm 1.0. Both modalities are projected into a shared latent space of dimension $d_{\text{proj}}=256$ followed by $\ell_2$ normalization. The learnable sigmoid scale and bias in the pairwise matching objective are initialized as $\alpha_0=10.0$ and $\beta_0=-10.0$, respectively; the scale is clamped to a stable optimization range during training.
    
    \paragraph{Dual-stream implementation.}
    For each modality $m\in\{s,y\}$, the encoder $\Phi_m$ returns a global feature $\mathbf{g}_m$ and token-level features $\mathbf{T}_m$. The global branch applies a linear projection head $\mathrm{Proj}_g(\cdot)$ to $\mathbf{g}_m$ to obtain the stable retrieval embedding $\mathbf{z}_m$. The fine-grained branch applies a learnable query-based Multi-Head Attention Pooling (MAP) operator over $\mathbf{T}_m$, followed by an independent linear head $\mathrm{Proj}_l(\cdot)$, yielding the localized embedding $\tilde{\mathbf{z}}_m$. This realizes the dual-stream representation
    \begin{equation}
    \mathbf{z}_m = \mathrm{Proj}_g(\mathbf{g}_m), \qquad
    \tilde{\mathbf{z}}_m = \mathrm{Proj}_l\big(\mathrm{MAP}(\mathbf{T}_m)\big),
    \end{equation}
    exactly as described in the main paper.
    
    \paragraph{Self-distillation in the pre-projection space.}
    The localized branch is sharper but less stable during early training. In implementation, self-distillation is therefore applied in the pre-projection feature space. Here, the pre-projection global feature $\mathbf{h}_{m,i}$ corresponds to the encoder global output $\mathbf{g}_{m,i}$, while the localized pre-projection feature is obtained by applying MAP to the token-level features, i.e.,
    \begin{equation}
    \mathbf{h}_{m,i} = \mathbf{g}_{m,i}, \qquad
    \tilde{\mathbf{h}}_{m,i} = \mathrm{MAP}(\mathbf{T}_{m,i}).
    \end{equation}
    A shallow predictor $\psi_m$ then maps $\tilde{\mathbf{h}}_{m,i}$ toward the stop-gradient target $\mathbf{h}_{m,i}$. This stabilizes the student-like localized branch without directly perturbing the global retrieval logits used at inference.
    
    \paragraph{Absorber-aware auxiliary regularization.}
    The absorber-aware term $\mathcal{L}_{\mathrm{absorber}}$ is computed only on the global-stream logits. All off-diagonal same-absorber pairs within a batch are treated as auxiliary positives, although they remain negatives under the exact-pair sigmoid matching objective. This term does not by itself enforce within-absorber separation. Instead, it softens the coarse absorber-dominated partition induced by exact-pair supervision alone, thereby contracting the absorber-level manifold while leaving sample-level discrimination to be resolved jointly by $\mathcal{L}_{\mathrm{global}}$, $\mathcal{L}_{\mathrm{fg}}$, and $\mathcal{L}_{\mathrm{distill}}$. In other words, $\mathcal{L}_{\mathrm{absorber}}$ acts as a coarse manifold-shaping prior rather than as an independent fine-grained discriminative objective.
    
    \paragraph{Training-only regularizers and inference path.}
    To clarify the exact role and computational footprint of each component, Table~\ref{tab:xaslip_components} summarizes the four terms in the full XASLip objective. In the main configuration, the corresponding weights are $\lambda_{\mathrm{fg}}=0.5$, $\lambda_{\mathrm{dist}}=0.1$, and $\lambda_{\mathrm{abs}}=0.001$. Importantly, the fine-grained matching branch and the self-distillation predictors are used only during training to shape the latent manifold. During inference and downstream retrieval, only the stable global embeddings $(\mathbf{z}_s,\mathbf{z}_y)$ are retained, exactly matching the inference protocol. As a result, the train-time localized branch introduces no additional inference overhead for retrieval, forward prediction, or inverse generation.
    
    \begin{table}[h]
    \centering
    \small
    \caption{Summary of XASLip objective components. The localized branch and predictor networks are used only during training to improve fine-grained alignment without increasing inference-time cost.}
    \label{tab:xaslip_components}
    \begin{tabular}{llcc}
        \toprule
        Loss Component & Representation Branch & Weight & Required at Inference \\
        \midrule
        Global Matching ($\mathcal{L}_{\mathrm{global}}$) & Global ($\mathbf{z}_s, \mathbf{z}_y$) & $1.0$ & \textbf{Yes} \\
        Fine-grained Matching ($\mathcal{L}_{\mathrm{fg}}$) & Localized ($\tilde{\mathbf{z}}_s, \tilde{\mathbf{z}}_y$) & $0.5$ & No \\
        Self-Distillation ($\mathcal{L}_{\mathrm{distill}}$) & Local $\rightarrow$ Global & $0.1$ & No \\
        Absorber Regularizer ($\mathcal{L}_{\mathrm{absorber}}$) & Global ($\mathbf{z}_s, \mathbf{z}_y$) & $0.001$ & No \\
        \bottomrule
    \end{tabular}
    \end{table}
    
    \subsection{Forward Prediction Module}
    \label{app:forward_train}
    
    This subsection provides the implementation details of the forward prediction module described in Sec.~4.4. Consistent with the main text, the forward stage operates on top of the frozen aligned latent space and integrates retrieval-augmented decoding, decomposed physical prediction, and latent manifold consistency.
    
    The forward module is trained after freezing the aligned latent interface. The main configuration uses batch size 256, learning rate $5\times10^{-4}$, weight decay $10^{-4}$, and 5\% linear warm-up followed by cosine decay.
    
    \paragraph{Retrieval-augmented cross-attention decoding.}
    Following Sec.~4.4, we perform $K=6$ nearest-neighbor retrieval in the shared latent space $\mathcal{Z}$ using cosine similarity between the structural query $\mathbf{z}_s$ and training spectral embeddings. 
    
    Moreover, we augment the retrieved memory tokens with lightweight intensity statistics from the training spectra, including baseline bias and RMS scale, which encode physical priors on the background absorption level and overall absorption magnitude. These statistics are derived from retrieved training neighbors rather than the target sample itself, and therefore provide approximate priors without leaking ground-truth information. The final intensity parameters are still predicted by the decoder.
    
    The retrieved tokens are consumed by a Transformer-based cross-attention decoder with 6 layers, 8 attention heads, and dropout 0.1. The retrieval residual gate is zero-initialized, ensuring that the decoder initially behaves as a purely parametric predictor and only gradually incorporates retrieved information as auxiliary guidance.
    
    \paragraph{Decomposed physical prediction head.}
    Consistent with the decomposed formulation in the main text, the forward head predicts
    \begin{equation}
    \hat{\mathbf{I}}=\hat{g}\cdot \hat{\mathbf{I}}_{\text{shape}}+\hat{b},
    \end{equation}
    where the morphology component is explicitly normalized. Let $\hat{\mathbf{I}}_{\text{raw}}$ denote the raw decoder output. The shape branch is defined as
    \begin{equation}
    \hat{\mathbf{I}}_{\text{shape}}
    =
    \frac{\hat{\mathbf{I}}_{\text{raw}}-\mathrm{mean}(\hat{\mathbf{I}}_{\text{raw}})}
    {\mathrm{RMS}\!\big(\hat{\mathbf{I}}_{\text{raw}}-\mathrm{mean}(\hat{\mathbf{I}}_{\text{raw}})\big)+\varepsilon},
    \end{equation}
    where $\mathrm{RMS}(\mathbf{x}) = \sqrt{\frac{1}{T}\sum_t x_t^2}$ denotes the root-mean-square operator. The gain $\hat g$ is constrained to be positive via a softplus head, and $\hat b$ is a scalar baseline bias. 
    
    This decomposition matches the formulation in Sec.~4.4 and ensures identifiability by explicitly separating morphology, global scale, and baseline level, enabling stable learning of absolute (non-normalized) spectra.
    
    \paragraph{Latent manifold consistency.}
    Let $\hat{\mathbf{Y}}=[\hat{\tilde{E}}_0,\hat{\mathbf{I}}]$ denote the predicted standardized spectrum. We re-encode $\hat{\mathbf{Y}}$ through the frozen spectral encoder to obtain $\hat{\mathbf{z}}_y=\Phi_y(\hat{\mathbf{Y}})$. We then enforce alignment with both the structural query and the ground-truth embedding:
    \begin{equation}
    \mathcal{L}_{\text{align}} = \gamma_1 \big(1 - \cos(\hat{\mathbf{z}}_y, \mathbf{z}_s)\big) + \gamma_2 \big(1 - \cos(\hat{\mathbf{z}}_y, \mathbf{z}_y)\big).
    \end{equation}
    This provides a stable semantic constraint without introducing representation drift.
    
    \paragraph{Auxiliary local structure channel.}
    In addition to the main structural encoder $\Phi_s$, we retain one auxiliary locality-sensitive geometric side channel in the forward implementation. This module is not introduced as a separate methodological contribution; rather, it is a fixed implementation detail shared by all Uni-XAS forward variants. The forward ablations reported in Table~4 (Part II) of the main paper therefore compare retrieval augmentation, decomposed physical prediction, and latent manifold consistency under the same local-geometry side channel, so that the reported gains are not attributable to a changing structural encoder budget.
    
    \paragraph{Forward training objective.}
    The forward loss follows the decomposed supervision strategy. We apply MSE to the normalized anchor $\hat{\tilde{E}}_0$, and independently supervise the shape, gain, and bias components against their ground-truth counterparts derived via the same normalization procedure. In the main configuration, the loss weights are 0.02 (shape), 0.05 (gain), and 0.05 (bias). The latent consistency weights are set to $\gamma_1=0.1$ and $\gamma_2=0.1$.
    
    \subsection{Inverse PR-Flow Generator}
    \label{app:inverse_train}
    
    This subsection details the concrete implementation of the inverse generator introduced in Sec.~4.5. It does not define a different inverse problem from the main paper. Instead, it specifies the latent-conditioning module, decoder instantiation, training stabilizers, and sampling settings used to realize the composition-conditional PR-Flow model under the benchmark setting used in this work.
    
    The inverse generator is trained after freezing the aligned encoders $\Phi_s$ and $\Phi_y$, following the staged training schedule in Sec.~4.5. The main inverse configuration uses batch size 256, 500 epochs, learning rate $5\times10^{-4}$, weight decay $10^{-4}$, gradient clipping at 1.0, and 6\% linear warm-up. Validation is run every epoch. During training, we use a reduced-cost sampling configuration for fast validation; after epoch 400, candidate checkpoints are re-evaluated using the full 150-step sampling setup before final checkpoint selection.
    
    \paragraph{Conditioning via latent prototypes.}
    As described in Sec.~4.5, given a query spectrum $\mathbf{Y}$, we first compute the spectral embedding $\mathbf{z}_y=\Phi_y(\mathbf{Y})$ in the frozen shared latent space. Cross-modal retrieval is then performed against a bank of training-structure embeddings to obtain structural prototypes. During training, exact sample-ID matches are masked from the retrieval bank; at inference time, the retrieval bank is restricted to the training split only, consistent with the split and retrieval rules summarized in Sections~A--C.
    
    In the main inverse configuration, retrieval is enabled with top-$K=2$ structural neighbors. If the mean similarity of the retrieved set falls below a confidence threshold of 0.20, the retrieved memory is replaced by a direct fallback formed from the spectral embedding itself. Retrieved prototypes are then reweighted by a softmax with temperature 0.70 before being passed to the latent-conditioning module. This design keeps the conditioning mechanism robust when retrieval quality is low while retaining a uniform decoder interface.
    
    The latent-conditioning module uses learned latent tokens of length 16, hidden width $d_{\text{latent}}=256$, 4 refinement layers, 8 attention heads, and dropout 0.1. Projected retrieved structure embeddings and the projected spectrum embedding are concatenated into a conditioning memory, and the latent tokens are refined through alternating self-attention and cross-attention blocks. These settings instantiate the latent-prototype conditioning mechanism described in Sec.~4.5; they are implementation choices rather than additional methodological assumptions.
    
    \paragraph{Mean-centered coordinate convention.}
    Sec.~4.5 formulates PR-Flow on mean-centered coordinates in order to remove global translation and focus learning on relative geometry. For non-padded (valid) atoms, we use
    \begin{equation}
        \bar{\mathbf{X}}
        =
        \mathbf{X}
        -
        \frac{1}{N_{\mathrm{valid}}}
        \sum_{n \in \mathrm{valid}} \mathbf{x}_n.
    \end{equation}
    Although each local environment is physically cropped around the absorber, inverse training is performed in this mean-centered frame. The same centering operation is applied to both target coordinates and sampled Gaussian noise before constructing the interpolation path. This makes the supervised flow target consistent with Eq.~(11) of the main text and removes residual global translation ambiguity at both endpoints.
    
    \paragraph{$E(3)$-equivariant decoder instantiation.}
     Sec.~4.5 describes the inverse decoder abstractly as predicting the flow field through lightweight $E(3)$-equivariant updates constructed from invariant scalar weights and relative displacement vectors. Here we specify the concrete realization used in our experiments.
    
    The inverse decoder uses 8 Transformer blocks with hidden width 256 and 8 attention heads. Pairwise distances are expanded with 64 Gaussian radial bases up to $10.0$\,\AA. The decoder follows the design principle stated in the main text: invariant quantities are used only to predict scalar coefficients, while vector updates are formed exclusively from relative displacement vectors. For attention head $h$, the scalar attention logits are
    \begin{equation}
        \ell_{ij}^{(h)}
        =
        \frac{\langle \mathbf{q}_i^{(h)}, \mathbf{k}_j^{(h)} \rangle}{\sqrt{d_h}}
        -
        \|\mathbf{x}_j-\mathbf{x}_i\|_2^2,
    \end{equation}
    followed by normalized weights
    \begin{equation}
        a_{ij}^{(h)}=\mathrm{softmax}_j(\ell_{ij}^{(h)}).
    \end{equation}
    The per-head equivariant message is
    \begin{equation}
        \mathbf{m}_i^{(h)}=\sum_j a_{ij}^{(h)}(\mathbf{x}_j-\mathbf{x}_i),
    \end{equation}
    and the final coordinate update is
    \begin{equation}
        \Delta \mathbf{x}_i=\sum_h \gamma_i^{(h)}\mathbf{m}_i^{(h)},
    \end{equation}
    where $\gamma_i^{(h)}$ are scalar mixing coefficients predicted from invariant node features. This is the concrete Transformer-style realization of the lightweight $E(3)$-equivariant update rule described in Sec.~4.5. Since pairwise distances are invariant and relative displacements transform equivariantly under rigid motions, the resulting coordinate update is $E(3)$-equivariant at the decoder-update level.
    
    \paragraph{Permutation-Rectified Flow Matching in practice.}
     Sec.~4.5 defines PR-Flow as a type-wise optimal-transport coupling between mean-centered Gaussian noise and mean-centered target coordinates. In practice, we compute this coupling independently within each identical-element group using Hungarian matching, which exactly solves the corresponding discrete assignment problem for the squared Euclidean cost used in Eq.~(10). When the absorber atom is present as a distinct valid anchor, it is held fixed and excluded from permutation.
    
    This preprocessing step is gradient-free and is used only to define permutation-consistent training targets. After rectification, we use the straight-line path
    \begin{equation}
        \bar{\mathbf{X}}_t=(1-t)\bar{\mathbf{X}}_0+t\bar{\mathbf{X}}_1^{\mathrm{aligned}},
        \qquad t\in[0,1],
    \end{equation}
    whose target velocity is constant:
    \begin{equation}
        \frac{d\bar{\mathbf{X}}_t}{dt}
        =
        \bar{\mathbf{X}}_1^{\mathrm{aligned}}-\bar{\mathbf{X}}_0.
    \end{equation}
    In this formulation, PR-Flow resolves same-type permutation ambiguity at the target-construction level, while mean-centering removes global translation ambiguity. Consistent with the main text, this formulation does not impose an explicit rotation-canonical frame.
    
    \paragraph{Core flow objective and auxiliary stabilizer.}
    With the above notation, the decoder predicts a conditional vector field
    \[
    \mathbf{v}_{\theta}(\bar{\mathbf{X}}_t,t,\mathbf{Z}_{\mathrm{cond}})
    \]
    to regress the constant target velocity induced by the rectified interpolation path. The defining generative objective is the flow-matching loss in Eq.~(12) of the main text:
    \begin{equation}
        \mathcal{L}_{\mathrm{flow}}
        =
        \mathbb{E}_{t,\bar{\mathbf{X}}_0,\bar{\mathbf{X}}_1}
        \left\|
        \mathbf{v}_{\theta}(\bar{\mathbf{X}}_t,t,\mathbf{Z}_{\mathrm{cond}})
        -
        \left(\bar{\mathbf{X}}_1^{\mathrm{aligned}}-\bar{\mathbf{X}}_0\right)
        \right\|^2.
    \end{equation}
    
    To match the implementation used for the reported checkpoints, we additionally allow small auxiliary stabilizers during training. In the main inverse configuration, the direct coordinate reconstruction auxiliary term is disabled ($\lambda_{x_1}=0$), while a small absorber-centered radial regularizer is enabled with weight $\lambda_{\mathrm{rad}}=0.02$. This auxiliary term regularizes distances to the absorber and is used only as a training stabilizer; it does not change the inverse problem definition, the PR-Flow path construction, or the evaluation metrics reported in the main paper.
    
    \paragraph{Design rationale.}
    The inverse design follows the decomposition in Sec.~4.5. In our benchmark setting, atom types and node count are provided, so the remaining challenge is stable spectrum-conditioned recovery of relative 3D geometry. Within this setting, mean-centering removes global translation ambiguity, PR-Flow resolves permutation ambiguity among atoms of the same type, and the decoder-level $E(3)$-equivariant updates preserve geometric sensitivity without requiring heavier higher-order equivariant parameterizations. This yields a scalable implementation of the composition-conditional coordinate generator defined in the main text.
    
    \paragraph{Classifier-free guidance and sampling.}
    The main text introduces inverse inference via the idealized deterministic probability-flow ODE. Here we provide the exact sampling settings used in our experiments. During training, conditioning is dropped with probability 0.10 to enable classifier-free guidance. During sampling, classifier-free guidance uses scale 1.50 with a linear temporal ramp. The base trajectory is solved with Euler integration over 150 steps for full evaluation, while reduced-step sampling is used for fast validation during training. In practice, a stochastic exploration term with scale 0.03 is injected and annealed with $(1-t)^{0.5}$; this acts as a heuristic implementation stabilizer beyond the idealized ODE description to prevent local geometric collapse, which is a common practice in empirical 3D generation.
    
    Throughout sampling, coordinates are repeatedly re-centered over valid atoms to remain consistent with the mean-centered training formulation. Generated coordinates are also clipped to a maximum per-atom norm of $10.0$\,\AA\ for numerical stability. These are sampling-time implementation details for the model defined in Sec.~4.5.
    
\begin{table}[t]
    \centering
    \footnotesize % 这里改成 \footnotesize 会比 \small 更精致一点
    \caption{Main architecture settings for Uni-XAS.}
    \label{tab:app_arch}
    \begin{tabular}{ll}
        \toprule
        Component & Setting \\
        \midrule
        Structure encoder & Uni-Mol Large, 15 layers, $d=512$, FFN 2048, 64 heads \\
        Spectral encoder & sequence length 129, $d_{\text{model}}=128$, $d_{\text{ff}}=128$, 3 layers \\
        Shared projection & latent dimension 256 \\
        Forward decoder & top-$K=6$, 6 cross-attention layers, 8 heads, dropout 0.1 \\
        Inverse retrieval & top-$K=2$, fallback threshold 0.20, temperature 0.70 \\
        Inverse latent module & length 16, $d_{\text{latent}}=256$, 4 layers, 8 heads \\
        Inverse flow decoder & 8 layers, 8 heads, 64 RBF bases, $r_{\max}=10.0$\,\AA \\
        Sampling & Euler ODE, $T_s=150$, $S=5$ hypotheses \\
        \bottomrule
    \end{tabular}
\end{table}

\begin{table}[t]
    \centering
    \footnotesize
    \caption{Stage-wise training hyperparameters for the main Uni-XAS pipeline.}
    \label{tab:app_train}
    \begin{tabular}{lcccc}
        \toprule
        Stage & Batch & Epochs & LR & Warm-up \\
        \midrule
        XASLip alignment & 128 & 500 & $1\times10^{-3}$ & -- \\
        Forward prediction & 256 & 500 & $5\times10^{-4}$ & 5\% \\
        Inverse generation & 256 & 500 & $5\times10^{-4}$ & 6\% \\
        \bottomrule
    \end{tabular}
\end{table}

% ================================================================
% E  Baselines
% ================================================================
\section{Baseline Provenance, Adaptation Scope, and Final Configurations}
\label{app:baselines}

    Several baselines considered in this work were not originally designed for our unified target representation or for spectrum-conditioned 3D local-structure generation. To describe the comparison clearly, we distinguish three categories of changes:
    
    \begin{itemize}[leftmargin=*,nosep]
        \item \textbf{Required interface changes:} minimal changes needed to run a method on our benchmark input/output format, such as replacing the output dimensionality, adding an XANES conditioning branch, or adapting a dataloader to absorber-centered local graphs.
        \item \textbf{Reconstruction choices:} changes required when official source code is unavailable and the method must be reconstructed from the published paper description.
        \item \textbf{Limited hyperparameter tuning:} a constrained search over a small set of optimizer and regularization hyperparameters under a fixed validation budget.
    \end{itemize}
    
    The resulting baselines should be read as controls evaluated under the same split, target definition, and metrics used in Section~5.1 of the main paper, rather than as authoritative reproductions of each source paper in its native dataset. Whenever official code is unavailable, we reconstruct only the minimum source-consistent architecture required to operate under the shared target $\mathbf{Y}=[\tilde{E}_0,\mathbf{I}]$ or the shared inverse task $p(\mathbf{X}\mid\mathbf{Y},\mathcal{V})$, and we report the resulting numbers conservatively. Under this setup, the fairest direct parametric comparison in Table~2 and Table~3 is against Uni-XAS (w/o RAG), while the gap from Uni-XAS (w/o RAG) to Uni-XAS (Full) quantifies the additional non-parametric benefit of retrieval conditioning within the same backbone family.

\begin{table*}[t]
    \centering
    \small
    \caption{Baseline provenance and adaptation scope. Availability basis describes what was available to our study when building the reference implementation under the same benchmark setting. To isolate the specific performance gains of our retrieval-augmented design, all external baselines are kept as purely parametric models without access to the retrieval memory bank.}
    \label{tab:baseline_provenance}
    \resizebox{\textwidth}{!}{%
    \begin{tabular}{L{1.7cm} L{2.8cm} L{2.8cm} L{2.4cm} L{7.0cm}}
        \toprule
        Method & Original formulation & Availability basis in this study & Native output / task & Required changes under our benchmark \\
        \midrule
        CLIP & generic dual-encoder contrastive alignment & objective re-instantiated in-house under shared XAS encoders & paired cross-modal retrieval objective & no architectural change beyond replacing image/text encoders with the shared XAS encoders \\
        SigLIP & pairwise sigmoid contrastive alignment & objective re-instantiated in-house under shared XAS encoders & paired cross-modal retrieval objective & same as CLIP; only the loss formulation differs \\
        SigLIP2 & dual-stream self-distilled contrastive alignment & objective re-instantiated in-house under shared XAS encoders & paired cross-modal retrieval objective & same shared encoders; MAP branch and self-distillation follow the SigLIP2-style objective design \\
        \midrule
        OmniXAS & structure-to-spectrum XAS surrogate using M3GNet features and XASBlock & method description plus released design pattern & normalized spectral regression & output expanded to 129-D target; common $E_0$ normalization adopted; feature extraction path converted to end-to-end GPU training to remain feasible at benchmark scale \\
        XAStruct & graph encoder + SGMLP head for XAS prediction & paper description only; reconstructed in-house & spectral regression / descriptor prediction & forward branch reconstructed; output expanded to 129-D anchored absolute-spectrum target; common dataloader and evaluation path adopted \\
        CGXAS & crystal graph XANES prediction with geometric encodings & paper description only; reconstructed in-house & spectral regression & graph pipeline reconstructed; output expanded to 129-D anchored absolute-spectrum target; same physical denormalization and evaluation path adopted \\
        \midrule
        EDM & $E(3)$-equivariant diffusion for 3D molecular generation & adapted from released molecular generation code & unconditional / property-conditioned 3D generation & XANES condition encoder added; conditioning concatenated outside equivariant coordinate updates; unified inverse evaluator adopted \\
        FlowMol3 & flow-matching molecular generator with GVP dynamics & adapted from released flow-matching code & conditional 3D molecular generation & XANES condition encoder added; graph-level conditioning injected into node scalars; unified inverse evaluator adopted \\
        DiffCSP-PP & diffusion model for crystal structure prediction & adapted from released crystal generation code & crystal denoising / structure prediction & XANES condition encoder added via time-condition fusion; local-structure dataloader and unified inverse evaluator adopted \\
        IDFlow & IPA-based SE(3) flow model & adapted from released SE(3) flow code & energy-conditioned 3D generation & XANES encoder added; conditioning concatenated into node initialization while keeping original IPA backbone structure \\
        \bottomrule
    \end{tabular}
    }
\end{table*}

\begin{table*}[t]
    \centering
    \small
    \caption{Tuning budget used to keep model selection limited and easy to inspect. The intention is not exhaustive architecture search, but a fixed-budget sweep over a small number of optimizer and regularization choices while keeping each model family close to its source design.}
    \label{tab:tuning_budget}
    \resizebox{\textwidth}{!}{%
    \begin{tabular}{L{2.8cm} L{4.3cm} L{4.4cm} L{3.2cm} L{3.0cm}}
        \toprule
        Family & Shared constraints & Primary tuned hyperparameters & Search policy & Validation selector \\
        \midrule
        Retrieval references and XASLip &
        same train/val split; same structure and spectral encoders; same projection dimension; same batch construction &
        learning rate, weight decay, contrastive temperature initialization, and for XASLip the loss weights $\lambda_{\mathrm{fg}}$ and $\lambda_{\mathrm{abs}}$ &
        small fixed grid; backbone widths and layer counts are not re-designed per loss &
        average bidirectional MRR \\
        \midrule
        Forward baselines &
        same 129-D target; same $E_0$ normalization table; same data split; no retrieval memory &
        learning rate, weight decay, batch size (memory permitting), dropout, and minor head-width choices kept within source-consistent ranges &
        small fixed budget over optimizer and regularization settings; no large architecture sweep &
        the same validation monitor implemented by each baseline (primarily validation loss; intensity $R^2$ additionally tracked where available) \\
        \midrule
        Uni-XAS forward &
        frozen aligned encoders; same target and train/val split; train-bank-only retrieval &
        learning rate, latent alignment weight $\gamma$, retrieval top-$K$, same-element retrieval bias, and auxiliary regularizer strengths &
        constrained search around the main architecture; zero-gated retrieval and decomposed head are kept fixed &
        validation intensity $R^2$ with loss and $E_0$ error archived jointly \\
        \midrule
        External inverse baselines &
        same local-structure dataset; same node-count/composition-controlled setting; same unified evaluator; no retrieval bank &
        learning rate, batch size, conditioning dropout, and condition-encoder width matched to each native backbone scale &
        small family-specific search while retaining native decoder class and native symmetry constraints &
        baseline-specific validation monitor; final selected checkpoint is evaluated only once with the common inverse metrics \\
        \midrule
        Uni-XAS inverse &
        frozen aligned encoders; train-bank-only retrieval; same full-eval script and $S=5$ hypotheses &
        learning rate, retrieval top-$K$, retrieval temperature, CFG scale, and conditioning dropout; PR-Flow and latent conditioning remain enabled in the full model &
        constrained search around the main retrieval-conditioned inverse design; no alternative decoder family is introduced &
        validation OT-RMSD$_{\mathrm{BoS}}$; candidate checkpoints re-checked by full validation after the epoch threshold \\
        \bottomrule
    \end{tabular}
    }
\end{table*}

\subsection{Forward Baseline Adaptation}
\label{app:forward_baselines}

\subsubsection{Common Absolute-Spectrum Setup}
\label{app:forward_protocol}

Prior forward XAS surrogates typically predict normalized spectral shapes without explicitly modeling the absolute edge anchor. To evaluate them under the same anchored absolute-spectrum setting, we apply the following changes:

\begin{enumerate}[leftmargin=*,nosep]
    \item \textbf{Target reformulation.} Each baseline is retrained to regress the full 129-dimensional target $\mathbf{Y}=[\tilde{E}_0,\mathbf{I}]$.
    \item \textbf{Shared $E_0$ normalization.} All methods use the same absorber-specific normalization and denormalization lookup table.
    \item \textbf{Common evaluation rule.} Reported $E_0$ errors are always computed after denormalization into eV.
        \item \textbf{Common data split and dataloader contract.} All baselines use the same absorber-centered local graph definition and the same training / held-out evaluation split, with the configs retaining the legacy \texttt{valid\_lmdb} name for the held-out evaluation partition.
\end{enumerate}

These modifications are necessary to evaluate all forward models under the same setting; they should not be read as claiming equivalence to the original papers' native data settings.

    \subsubsection{OmniXAS Adaptation}
    \label{app:omnixas}
    
    OmniXAS~\cite{kharel2025omnixas} follows a structure-to-spectrum pipeline built around M3GNet-derived atom features and an MLP-style XASBlock predictor. In its original formulation, the method operates on normalized spectra and relies on an offline feature workflow.
    
    \paragraph{Required changes.}
    The output dimension is expanded to 129 to model the anchored absolute-spectrum target. The common absorber-specific $E_0$ normalization table is used, and all evaluation uses the same normalization and metric code. To keep training feasible at the scale of our benchmark, we replace the offline feature-cache path with on-the-fly graph batching and end-to-end GPU execution. This preserves the high-level structure-encoder-to-spectrum-head design of OmniXAS while avoiding a prohibitive preprocessing bottleneck at our data scale.
    
    \paragraph{Selected configuration.}
    The adapted OmniXAS reference uses an M3GNet backbone with an XASBlock head of hidden dimensions $[512,512,256]$. The backbone is optimized with a reduced learning rate multiplier relative to the head in order to retain useful geometric priors while allowing the new anchored target head to adapt. Validation loss is used as the primary checkpoint monitor.

    \subsubsection{XAStruct Adaptation}
    \label{app:xastruct}
    
    XAStruct~\cite{wang2025xastruct} employs a graph encoder followed by a gated SGMLP prediction head. No official code was available to our study, so this baseline is an in-house reimplementation based on the published paper description. We use XAStruct only as a forward-task reference in our study. Its inverse branch predicts structural descriptors rather than explicit 3D coordinates and therefore is not directly comparable to the coordinate-generation setting evaluated by OT-RMSD and RDF-L1 in Table~3.
    
    \paragraph{Required changes.}
    The forward branch is reconstructed for absorber-centered local graphs, and the SGMLP head output is expanded to 129 dimensions to predict $[\tilde{E}_0,\mathbf{I}]$. No separate $E_0$ branch is introduced, which keeps the baseline close to its original monolithic predictor design.

    \paragraph{Selected configuration.}
    The reconstructed XAStruct reference uses a graph encoder followed by a gated SGMLP-style prediction head under the absorber-centered local-graph protocol. Validation loss is used for checkpoint selection under the common forward-task evaluation setting.

    \subsubsection{CGXAS Adaptation}
    \label{app:cgxas}
    
    CGXAS~\cite{lin2026cgxas} combines crystal-graph message passing with an MLP decoder for XANES prediction. As with XAStruct, no official repository was available to our study; the baseline is therefore reconstructed from the paper description and adapted to our protocol.
    
    \paragraph{Required changes.}
    The graph pipeline is reformulated to ingest absorber-centered local structures instead of the original crystal-level setting, and the decoder output is expanded to 129 dimensions. The same shared $E_0$ normalization and evaluation path is used.
    
    \paragraph{Selected configuration.}
    The adapted CGXAS reference retains gated geometric message passing and uses an MLP head with positive-output activation under the same normalization and metric setup. Validation intensity $R^2$ is monitored together with validation loss, and both checkpoints are archived.
    
    \paragraph{Interpretation.}
    All three forward baselines process the unified 129-dimensional target through a single regression head, whereas Uni-XAS introduces an explicitly decomposed physical prediction head together with retrieval-augmented calibration. This comparison is meant to quantify the value of these additional inductive biases under the same split and metrics, rather than to claim that the source methods are intrinsically limited in their native tasks.

\subsection{Inverse Generative Baseline Adaptation}
\label{app:inverse_baselines}

    \subsubsection{Adaptation Principles}
    \label{app:inverse_principles}
    
    There is currently no established shared control suite for explicit XANES-conditioned local-structure generation under one common benchmark setting. While the broader inverse spectroscopy literature includes descriptor prediction, fitting-oriented recovery, and emerging direct generative approaches in narrower settings, these do not yet provide a baseline ecosystem with the same split and evaluator used here. We therefore build one control suite by adapting representative molecular and crystal 3D generative frameworks to the spectrum-conditioned setting.
    
    \paragraph{Concurrent work.}
    Concurrent to our study, Okubo et al.~\cite{okubo2026generative} reported direct near-edge-spectrum-to-3D coordination generation using an equivariant diffusion model in a materially narrower Si--O setting. We therefore do not claim to be the first work on generative inverse near-edge spectroscopy in an absolute sense. Our claim is narrower: to the best of our knowledge, this is the first large paired XAS benchmark that studies cross-modal retrieval, anchored absolute-spectrum prediction, and inverse local-structure generation under one split, target definition, and evaluator. Because the concurrent study does not provide an implementation and adopts a substantially different generative formulation (e.g., equivariant diffusion with task-specific design choices), a faithful reproduction under our setting would require additional modeling assumptions beyond straightforward re-implementation. This differs from the baselines included in our benchmark, which share comparable task definitions and can be evaluated with the same split and metrics. To avoid introducing potentially unfair or non-reproducible comparisons, we therefore discuss the concurrent work qualitatively rather than include it as a standardized control in Table~3.
    
    One further choice matters for fairness. Consistent with the inverse formulation in the main text, all inverse models are evaluated under the same composition-controlled setting: atom types and node count are provided by the benchmark, and the learning target is restricted to 3D coordinates. In other words, no external baseline is asked to infer composition or graph size. Whenever a source method originally supports joint generation of discrete and continuous variables, we disable that functionality and adapt it only to the coordinate-generation setting required here. This ensures that all compared methods solve the same conditional task $p(\mathbf{X}\mid\mathbf{Y},\mathcal{V})$ under the same evaluator.

    Three design choices are important:
    \begin{itemize}[leftmargin=*,nosep]
        \item all external inverse baselines are kept \textbf{purely parametric} and do not access the Uni-XAS retrieval bank;
        \item all models are evaluated under the same \textbf{composition-controlled setting}, where atom types and node count are given and only coordinates are generated;
        \item all methods are evaluated with the \textbf{same} OT-RMSD / RDF-L1 metrics, the same number of generated hypotheses, and the same held-out split.
    \end{itemize}
    
    We therefore treat these baselines as \emph{cross-domain adapted controls} rather than native XAS generative baselines. In particular, the principal apples-to-apples parametric comparison in Table~3 is between the external baselines and Uni-XAS (w/o RAG), while the gap from Uni-XAS (w/o RAG) to Uni-XAS (Full) quantifies the additional non-parametric benefit of retrieval conditioning within our own framework.

    \subsubsection{EDM Adaptation}
    \label{app:edm}
    
    EDM~\cite{hoogeboom2022edm} is an $E(3)$-equivariant diffusion model originally designed for 3D molecular generation.
    
    \paragraph{Required change.}
    We add a dedicated XANES condition encoder built from 1D convolutions and self-attention layers ($d_{\text{model}}=128$, 3 layers), followed by pooling and projection into a global conditioning vector. An absorber-type embedding is concatenated before fusion. The adapted EDM baseline is evaluated under the same composition-controlled setting as Uni-XAS inverse: atom types and node count are provided by the benchmark, and only coordinates are generated.
    
    \paragraph{Conditioning injection.}
    The conditioning vector is projected to the EGNN hidden size and concatenated with the time embedding and atom-type embedding before the input projection:
    \begin{equation}
        \mathbf{h}_i^{(0)}=\mathrm{Linear}\!\big([\mathbf{e}_{\mathrm{type},i}\,\|\,\mathbf{t}_{\mathrm{emb}}\,\|\,\mathbf{c}]\big).
    \end{equation}
    Because conditioning is injected only through invariant node features, the equivariance of coordinate updates is preserved.

    \subsubsection{FlowMol3 Adaptation}
    \label{app:flowmol3}
    
    FlowMol3~\cite{dunn2508flowmol3} is a flow-matching model that uses GVP-based vector field networks for 3D molecular generation.
    
    \paragraph{Required change.}
    We attach a spectrum encoder of the same scale as EDM to produce a graph-level conditioning vector. As with all inverse baselines in this study, the adapted FlowMol3 model is evaluated under the composition-controlled setting with known atom types and node count, so the generation target is restricted to coordinates only.
    
    \paragraph{Conditioning injection.}
    The spectral condition is broadcast to all nodes and concatenated with atom-type, charge, and time embeddings:
    \begin{equation}
        \mathbf{s}_i^{(0)}=\mathrm{ScalarEmbed}\!\big([\mathbf{e}_{\mathrm{atom},i}\,\|\,\mathbf{e}_{\mathrm{charge},i}\,\|\,\mathbf{t}_{\mathrm{emb}}\,\|\,\mathbf{c}]\big).
    \end{equation}
    The downstream GVP dynamics are otherwise retained.

    \subsubsection{DiffCSP-PP Adaptation}
    \label{app:diffcsp}
    
    DiffCSP-PP~\cite{jiao2024space} is a diffusion model for crystal structure prediction whose denoiser is globally conditioned through the time embedding.
    
    \paragraph{Required change.}
    We add a lightweight XANES encoder based on an MLP operating on the spectral vector and a validity mask, with absorber identity fused additively. The adapted DiffCSP-PP reference is likewise evaluated with known atom types and node count from the benchmark setting used here, and predicts only 3D coordinates.
    
    \paragraph{Conditioning injection.}
    The spectral condition is fused with the sinusoidal diffusion-time embedding:
    \begin{equation}
        \tilde{\mathbf{t}}=\mathrm{MLP}\!\big([\mathbf{t}_{\mathrm{emb}}\,\|\,\mathbf{c}]\big),
    \end{equation}
    and $\tilde{\mathbf{t}}$ replaces the original time embedding inside the CSPNet decoder. This provides global spectrum-conditioned modulation without altering the core denoiser family.

    \subsubsection{IDFlow Adaptation}
    \label{app:idflow}
    
    IDFlow~\cite{zhou2025energy} is an IPA-based flow model for 3D generation.
    
    \paragraph{Required change.}
    We add a Transformer-based XANES encoder ($d_{\text{model}}=128$, $d_{\text{ff}}=256$, 4 layers, 4 heads) operating on the intensity sequence and producing a 256-dimensional graph-level condition. Consistent with the shared inverse protocol, the adapted IDFlow model receives atom types and node count as given inputs and is evaluated only on coordinate generation.
    
    \paragraph{Conditioning injection.}
    The condition is broadcast and concatenated with positional features, diffuse-mask features, and the dual time embeddings used by the model:
    \begin{equation}
        \mathbf{h}_i^{(0)}=\mathrm{Linear}\!\big([\mathbf{p}_i\,\|\,m_i\,\|\,\mathbf{t}_{\mathrm{SO3}}\,\|\,\mathbf{t}_{\mathrm{R3}}\,\|\,\mathbf{c}]\big).
    \end{equation}
    The subsequent IPA stack is left structurally unchanged.

% ================================================================
% F  Diagnostics
% ================================================================
\section{Additional Experimental Diagnostics and Robustness Notes}
\label{app:diagnostics}

    \subsection{Constrained Cross-Modal Retrieval Diagnostics}
    \label{app:clip_extra}
    
The main paper reports global retrieval on the full held-out evaluation split. Because coarse inter-element separation can inflate retrieval scores in scientific settings, we additionally evaluate two stricter diagnostics that suppress trivial absorber-level shortcuts:
    
    \begin{itemize}[leftmargin=*,nosep]
        \item \textbf{Within-absorber:} for each query, the gallery is restricted to samples with the same central absorbing element;
        \item \textbf{Within-absorber + same node count:} the gallery is further restricted to samples with the same absorber and the same number of atoms in the local graph.
    \end{itemize}
    
    These settings force the model to rely more strongly on fine-grained spectral morphology and subtle 3D coordination variations. As shown in Table~\ref{tab:app_diagnostic_full}, the ranking pattern reported for global retrieval in Table~1 of the main paper and for the retrieval ablation in Table~4 (Part I) is preserved under both constrained diagnostics: Physics-Aware Encoding improves fine-grained sensitivity, and the full absorber-aware XASLip objective yields the strongest within-element discrimination. Together with the Fe-only similarity visualization in Figure~3B of the main paper, these results support the interpretation that the aligned latent space captures coordination-level structure rather than merely absorber identity.

\begin{table}[t]
    \centering
    \small
    \caption{Diagnostic retrieval ablations under constrained gallery settings. All metrics are percentages (\%). Best values are bolded.}
    \label{tab:app_diagnostic_full}
    \resizebox{\columnwidth}{!}{%
    \begin{tabular}{llcccccccc}
        \toprule
        \multirow{2}{*}{Setting} & \multirow{2}{*}{Method} & \multicolumn{4}{c}{Structure-to-Spectrum (S2P)} & \multicolumn{4}{c}{Spectrum-to-Structure (P2S)} \\
        \cmidrule(lr){3-6} \cmidrule(lr){7-10}
        & & R@1$\uparrow$ & R@5$\uparrow$ & R@10$\uparrow$ & MRR$\uparrow$ & R@1$\uparrow$ & R@5$\uparrow$ & R@10$\uparrow$ & MRR$\uparrow$ \\
        \midrule
        \multirow{4}{*}{\textbf{Within-Absorber}}
        & SigLIP2 & 33.05 & 69.97 & 81.47 & 49.26 & 32.93 & 70.52 & 82.37 & 49.40 \\
        \cmidrule{2-10}
        & Dual-Stream Baseline & 32.95 & 70.23 & 81.84 & 49.29 & 33.08 & 70.60 & 82.49 & 49.57 \\
        & + Physics-Aware Encoding & 36.14 & 72.65 & 83.39 & 52.15 & 36.52 & 73.71 & 84.55 & 52.77 \\
        & \textbf{+ Absorber-Aware (Full XASLip)} & \textbf{38.45} & \textbf{74.87} & \textbf{85.24} & \textbf{54.33} & \textbf{38.81} & \textbf{75.76} & \textbf{85.89} & \textbf{54.92} \\
        \midrule
        \multirow{4}{*}{\textbf{\shortstack[l]{Within-Absorber\\+ Same Node}}}
        & SigLIP2 & 65.67 & 93.79 & 96.91 & 77.86 & 66.95 & 94.22 & 97.11 & 78.84 \\
        \cmidrule{2-10}
        & Dual-Stream Baseline & 66.01 & 93.89 & 96.98 & 78.12 & 67.01 & 94.36 & 97.12 & 78.93 \\
        & + Physics-Aware Encoding & 67.61 & 94.23 & 97.03 & 79.22 & 68.78 & 94.77 & 97.29 & 80.14 \\
        & \textbf{+ Absorber-Aware (Full XASLip)} & \textbf{69.44} & \textbf{94.57} & \textbf{97.15} & \textbf{80.45} & \textbf{70.18} & \textbf{94.92} & \textbf{97.31} & \textbf{81.04} \\
        \bottomrule
    \end{tabular}
    }
\end{table}

    \subsection{Interpretation of Adapted Baseline Results}
    \label{app:diag_fairness}
    
    A recurring concern in emerging scientific benchmarks is that the authors may control both the benchmark design and the adapted baselines, especially when some source methods are unreleased or originally evaluated on non-comparable data. Our goal is therefore not to claim exact reproduction of every source paper’s reported numbers, but to provide a direct comparison under the same split, evaluator, and post-processing. We therefore emphasize the following:
    
    \begin{itemize}[leftmargin=*,nosep]
        \item the benchmark target representation is fixed \emph{before} baseline-specific adaptation;
        \item all adapted baselines use the same held-out split, the same evaluator, and the same post-processing scripts;
        \item all inverse baselines are denied access to the Uni-XAS retrieval bank by design;
        \item when official code is unavailable, we reconstruct only source-consistent reference implementations from the published descriptions and explicitly document the required changes;
        \item limited hyperparameter tuning is restricted to the fixed validation budget reported in Table~\ref{tab:tuning_budget}, and no held-out evaluation metric is used for model selection.
        \item whenever Uni-XAS Full is reported together with purely parametric external baselines, the corresponding Uni-XAS (w/o RAG) result should be read as the parametric reference under the same setup, while the Full-model gain quantifies the additional benefit of retrieval augmentation within the same backbone family.
    \end{itemize}
    
    These controls do not remove all ambiguity inherent to cross-domain adaptation, but they make the comparison easier to inspect and materially fairer. The empirical comparison should therefore be read as a comparison under the same split and metrics, rather than as a correction of the source papers’ original claims in their native settings.

% ================================================================
% G  Mathematics
% ================================================================
\section{Mathematical Definitions}
\label{app:math}

This section consolidates the full mathematical expressions used by the main paper for the physics-aware spectral encoder, the XASLip objective, and the physically meaningful inverse-generation metrics.

\subsection{Physics-Aware Spectral Encoder}
\label{app:math_spec}

The spectral encoder $\Phi_y$ processes the intensity signal together with its physically meaningful derivatives. Let $\mathbf{X}_{\mathrm{int}}\in\mathbb{R}^{L\times d_{\mathrm{model}}}$ denote the projected intensity tokens:
\begin{equation}
    \mathbf{X}_{\mathrm{int},t}
    =
    \mathbf{W}_{\mathrm{spec}}
    \begin{bmatrix}
        (\mathbf{I} * G_{\sigma})_t \\
        \Delta(\mathbf{I} * G_{\sigma})_t \\
        \Delta^2(\mathbf{I} * G_{\sigma})_t
    \end{bmatrix},
    \quad
    \forall t\in\{1,\dots,L\},
    \label{eq:spectral_input}
\end{equation}
where $G_{\sigma}$ is a Gaussian smoothing kernel, $\Delta$ and $\Delta^2$ denote first- and second-order finite-difference operators, and $\mathbf{W}_{\mathrm{spec}}\in\mathbb{R}^{d_{\mathrm{model}}\times 3}$ is a learnable projection matrix.

The normalized scalar edge anchor $\tilde{E}_0$ is embedded as a dedicated token
\[
\mathbf{x}_{\mathrm{anchor}}=\mathbf{w}_{E_0}\tilde{E}_0+\mathbf{b}_{E_0},
\]
and the initial sequence is
\begin{equation}
    \mathbf{X}^{(0)}=
    [\mathbf{x}_{\mathrm{anchor}},\mathbf{X}_{\mathrm{int},1},\dots,\mathbf{X}_{\mathrm{int},L}].
\end{equation}

The dynamic edge-onset location used by local pooling is defined as
\begin{equation}
    t^{\star}=\arg\max_t \Delta(\mathbf{I}*G_{\sigma})_t.
    \label{eq:onset_index}
\end{equation}

Given the final hidden states $\mathbf{X}^{(K)}=[\mathbf{x}_{\mathrm{anchor}}^{(K)},\mathbf{x}_1,\dots,\mathbf{x}_L]$, we extract a global spectral embedding together with an auxiliary edge-window summary:
\begin{equation}
    \mathbf{z}_y=\mathrm{MeanPool}(\mathbf{X}^{(K)}),
    \qquad
    \mathbf{z}^{\mathrm{edge}}_y=
    \mathrm{MeanPool}\big(\{\mathbf{x}_t \mid |t-t^{\star}|\leq w\}\big),
    \label{eq:onset_pooling}
\end{equation}
where $w$ is the pooling half-width. In the implemented XASLip objective, the fine-grained branch is obtained separately by applying MAP over the returned token-level spectral features rather than by replacing $\mathbf{z}_y$ with $\mathbf{z}^{\mathrm{edge}}_y$.

\subsection{XASLip Objective}
\label{app:math_xaslip}

\paragraph{Base sigmoid contrastive loss.}
Given $\ell_2$-normalized embeddings $\mathcal{A}=\{\mathbf{a}_i\}_{i=1}^{B}$ and $\mathcal{U}=\{\mathbf{u}_i\}_{i=1}^{B}$ from two modalities in a batch of size $B$, the pairwise sigmoid loss is
\begin{equation}
    \ell_{\mathrm{sig}}(\mathcal{A},\mathcal{U})
    =
    -\frac{1}{B^2}\sum_{i=1}^{B}\sum_{j=1}^{B}
    \log \sigma\!\Big(
        s_{ij}(\alpha\,\mathbf{a}_i^{\top}\mathbf{u}_j+\beta)
    \Big),
    \label{eq:sig_loss}
\end{equation}
where $\sigma(\cdot)$ is the sigmoid, $\alpha$ and $\beta$ are learnable scale and bias parameters, and
\[
s_{ij}=
\begin{cases}
1,& i=j,\\
-1,& i\neq j.
\end{cases}
\]

\paragraph{Global and localized matching.}
Using the pairwise sigmoid objective in Eq.~\ref{eq:sig_loss}, the global and localized alignment terms are
\begin{equation}
    \mathcal{L}_{\mathrm{global}}
    =
    \ell_{\mathrm{sig}}(\mathcal{Z}_s,\mathcal{Z}_y),
    \qquad
    \mathcal{L}_{\mathrm{fg}}
    =
    \ell_{\mathrm{sig}}(\tilde{\mathcal{Z}}_s,\tilde{\mathcal{Z}}_y).
\end{equation}
The first term governs the stable inference-time retrieval space, while the second acts as a sharper train-time fine-grained matching signal.

\paragraph{Cross-scale self-distillation.}
Let $\mathbf{h}_{s,i},\mathbf{h}_{y,i}$ denote the pre-projection global features and let $\tilde{\mathbf{h}}_{s,i},\tilde{\mathbf{h}}_{y,i}$ denote the corresponding localized features. To stabilize the more volatile localized branch during early training, we distill each localized feature toward its detached global counterpart using a cosine similarity constraint:
\begin{equation}
    \mathcal{L}_{\mathrm{distill}}
    =
    \frac{1}{2B}\sum_{i=1}^{B}
    \Big[
        2
        - \cos\big(\psi_s(\tilde{\mathbf{h}}_{s,i}), \mathrm{sg}(\mathbf{h}_{s,i})\big)
        - \cos\big(\psi_y(\tilde{\mathbf{h}}_{y,i}), \mathrm{sg}(\mathbf{h}_{y,i})\big)
    \Big],
\end{equation}
where $\psi_s$ and $\psi_y$ are shallow predictor networks and $\mathrm{sg}(\cdot)$ denotes the stop-gradient operation. This regularizer is applied in the pre-projection feature space, matching the implementation and preventing unnecessary interference with the global retrieval logits.

\paragraph{Absorber-aware auxiliary regularizer.}
Let $c_i$ denote the absorber species of sample $i$, and define the off-diagonal same-absorber pair set as
\[
\mathcal{P}_{\mathrm{abs}}=\{(i,j)\mid c_i=c_j,\ i\neq j\}.
\]
To reduce the incentive to solve alignment by coarse absorber identity alone, we apply a sigmoid-based absorber-aware auxiliary regularizer on the global-stream logits:
\begin{equation}
    \mathcal{L}_{\mathrm{absorber}}
    =
    -\frac{1}{|\mathcal{P}_{\mathrm{abs}}|}
    \sum_{(i,j)\in\mathcal{P}_{\mathrm{abs}}}
    \log \sigma \big(\alpha\,\mathbf{z}_{s,i}^{\top}\mathbf{z}_{y,j}+\beta\big).
\end{equation}
Unlike Eq.~\ref{eq:sig_loss}, this auxiliary term does not apply a sign flip to the pairs in $\mathcal{P}_{\mathrm{abs}}$; instead, it explicitly adds attraction to off-diagonal same-absorber pairs. Consequently, these pairs are simultaneously repelled by the exact-pair objective and attracted by the absorber-aware term. Importantly, $\mathcal{L}_{\mathrm{absorber}}$ alone does not impose any explicit within-absorber margin and, taken in isolation, would simply encourage same-absorber clustering. The fine-grained discrimination required for exact retrieval is instead maintained by $\mathcal{L}_{\mathrm{global}}$ and $\mathcal{L}_{\mathrm{fg}}$, with $\mathcal{L}_{\mathrm{distill}}$ stabilizing the localized branch. The desired coordination-sensitive within-absorber organization therefore comes from the joint XASLip objective, not from $\mathcal{L}_{\mathrm{absorber}}$ alone.

\paragraph{Overall objective.}
The final XASLip training objective integrates the global matching term, the localized fine-grained term, the cross-scale self-distillation term, and the absorber-aware auxiliary term:
\begin{equation}
    \mathcal{L}_{\mathrm{XASLip}}
    =
    \mathcal{L}_{\mathrm{global}}
    +
    \lambda_{\mathrm{fg}}\mathcal{L}_{\mathrm{fg}}
    +
    \lambda_{\mathrm{dist}}\mathcal{L}_{\mathrm{distill}}
    +
    \lambda_{\mathrm{abs}}\mathcal{L}_{\mathrm{absorber}}.
\end{equation}
with $\lambda_{\mathrm{fg}}=0.5$, $\lambda_{\mathrm{dist}}=0.1$, and $\lambda_{\mathrm{abs}}=0.001$ defining the main configuration.

\subsection{Evaluation Metrics}
    \label{app:math_metrics}
    
    \subsubsection{Element-wise Edge-Anchor Normalization and eV Conversion}
    \label{app:e0_norm}
    
    Because K-edge anchors vary substantially across absorber species, $E_0$ is normalized into an intra-element coordinate:
    \begin{equation}
        \tilde{E}_0
        =
        \frac{E_0-E_{\min}^{(c)}}{E_{\max}^{(c)}-E_{\min}^{(c)}},
    \end{equation}
    where $c$ is the absorber species and $(E_{\min}^{(c)},E_{\max}^{(c)})$ are element-specific bounds stored in \texttt{Energy\_Ele\_Norm.csv}.
    
    At inference, a predicted normalized anchor is mapped back into the physical energy domain:
    \begin{equation}
        \hat{E}_0
        =
        \hat{\tilde{E}}_0\big(E_{\max}^{(c)}-E_{\min}^{(c)}\big)+E_{\min}^{(c)}.
    \end{equation}
    
    The MAE in physical eV can be written as
    \begin{equation}
        \mathrm{MAE}(E_0)_{\mathrm{eV}}
        =
        \frac{1}{M}\sum_{i=1}^{M}
        \left|
            \hat{\tilde{E}}_{0,i}-\tilde{E}_{0,i}
        \right|
        \cdot
        \Delta E^{(c_i)},
    \end{equation}
    where $\Delta E^{(c_i)}=E_{\max}^{(c_i)}-E_{\min}^{(c_i)}$.
    
    \subsubsection{OT-RMSD}
    \label{app:ot_rmsd}
    
    Standard RMSD over-penalizes geometrically valid predictions when chemically identical atoms are permuted. We therefore evaluate inverse generation using type-consistent Optimal Transport RMSD (OT-RMSD)~\cite{grave2019unsupervised,xu2022geodiff,liang2024foundations}, which is invariant to both rigid-body motion and permutations among identical atom types.
    
    Consistent with the inverse formulation $p(\mathbf{X}\mid\mathbf{Y},\mathcal{V})$ in the main text, let $\Pi_{\mathrm{type}}$ denote the set of type-consistent permutations implied by $\mathcal{V}$, and let $(\mathbf{R} \in \mathrm{SO}(3), \boldsymbol{\tau})$ denote the rigid-body alignment. For generated coordinates $\hat{\mathbf{X}}\in\mathbb{R}^{N\times 3}$ and ground-truth coordinates $\mathbf{X}\in\mathbb{R}^{N\times 3}$, the OT-RMSD is defined as
    \begin{equation}
        \mathrm{OT\text{-}RMSD}(\hat{\mathbf{X}},\mathbf{X})
        =
        \min_{\substack{
            \pi \in \Pi_{\mathrm{type}}\\
            \mathbf{R}\in\mathrm{SO}(3)\\
            \boldsymbol{\tau}\in\mathbb{R}^{3}
        }}
        \left(
            \frac{1}{N}
            \sum_{i=1}^{N}
            \left\|
                \hat{\mathbf{X}}_{i}
                -
                \left(\mathbf{R}\mathbf{X}_{\pi(i)}+\boldsymbol{\tau}\right)
            \right\|_2^2
        \right)^{1/2}.
    \end{equation}
    
    In practice, this minimization is implemented by a fixed four-step alternation between type-wise linear assignment using the Hungarian algorithm and rigid alignment using the Kabsch algorithm, matching the provided evaluation script used for Table~3 of the main paper. This fixed-iteration implementation removes ambiguity in stopping behavior while preserving the same rigid-motion and same-type permutation invariances.
    
    For five generated hypotheses per held-out evaluation sample, we report both the expectation over generated hypotheses and the Best-of-5 score:
    \begin{align}
        \mathrm{OT\text{-}RMSD}_{\mathrm{Exp}}
        &=
        \frac{1}{M}\sum_{i=1}^{M}
        \frac{1}{5}\sum_{s=1}^{5}
        \mathrm{OT\text{-}RMSD}(\hat{\mathbf{X}}_{i}^{(s)},\mathbf{X}_{i}),
        \\
        \mathrm{Best\mbox{-}of\mbox{-}5\ OT\text{-}RMSD}
        &=
        \frac{1}{M}\sum_{i=1}^{M}
        \min_{1\le s\le 5}
        \mathrm{OT\text{-}RMSD}(\hat{\mathbf{X}}_{i}^{(s)},\mathbf{X}_{i}).
    \end{align}
    
    \subsubsection{RDF-L1}
    \label{app:rdf}
    
    To assess distributional realism beyond pointwise geometry, we compute the $L_1$ discrepancy between radial distribution functions (RDFs) of predicted and ground-truth structures. Using $B=64$ bins over $[0,r_{\max}]$ with $r_{\max}=8.0$\,\AA, the per-sample discrepancy is
    \begin{equation}
        \mathrm{RDF\text{-}L1}(\hat{\mathbf{X}}_i^{(s)},\mathbf{X}_i)
        =
        \frac{1}{B}\sum_{b=1}^{B}
        \left|
            \hat{g}_i^{(s)}(r_b)-g_i(r_b)
        \right|,
    \end{equation}
    where $\hat{g}_i^{(s)}(r_b)$ and $g_i(r_b)$ are the normalized RDF histogram values for the generated and ground-truth structures, respectively.
    
    RDF-L1 complements OT-RMSD: the latter measures atom-level geometric fidelity after type-consistent matching, whereas RDF-L1 measures whether the generated structure reproduces the correct radial-shell statistics even when the inverse mapping is one-to-many.

% ================================================================
% H  Extended Qualitative Visualizations
% ================================================================
\begin{figure*}[htbp]
    \centering
    \includegraphics[width=0.9\linewidth]{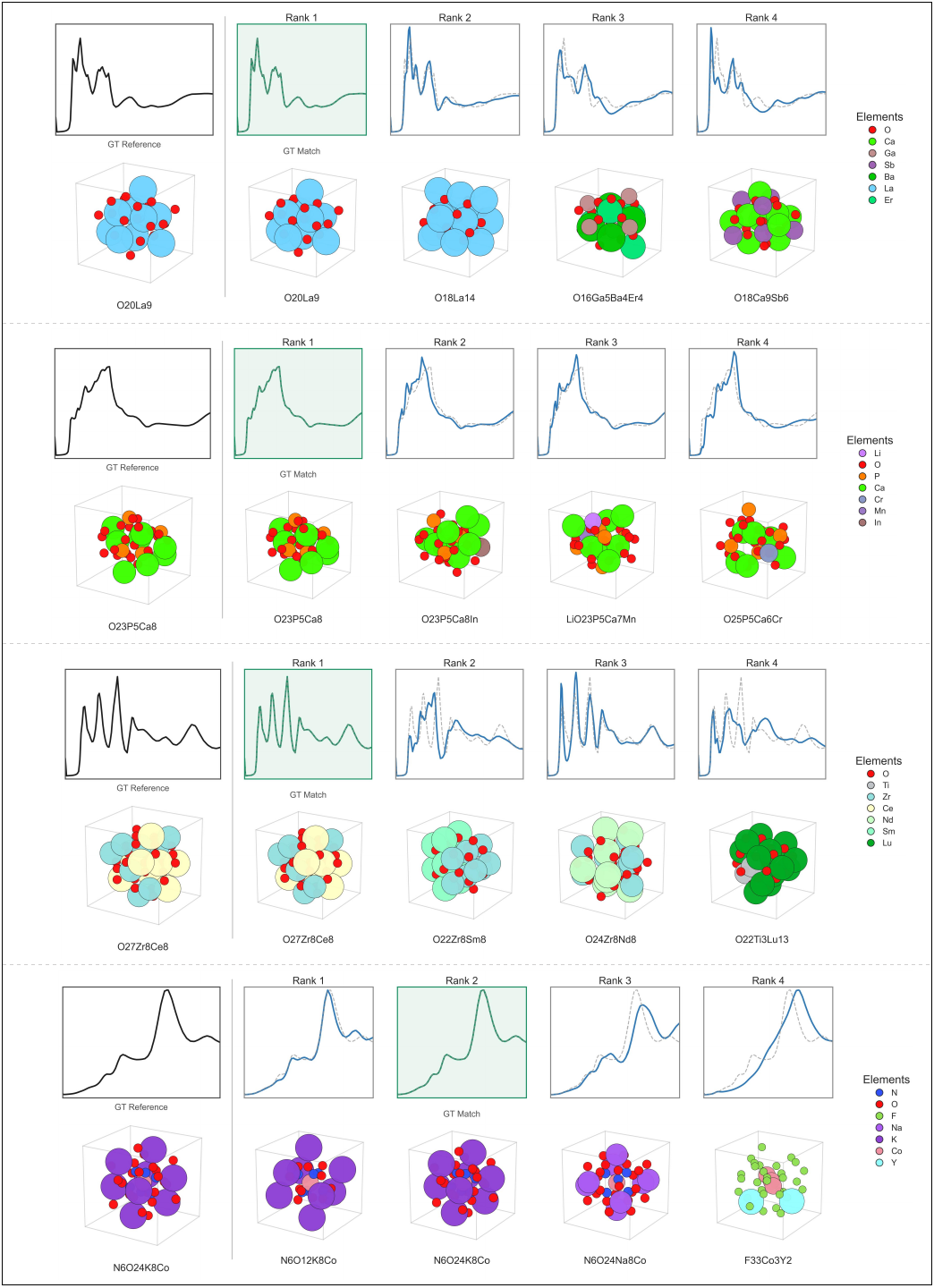} 
    
\end{figure*}
\begin{figure*}[htbp]
    \includegraphics[width=0.9\linewidth]{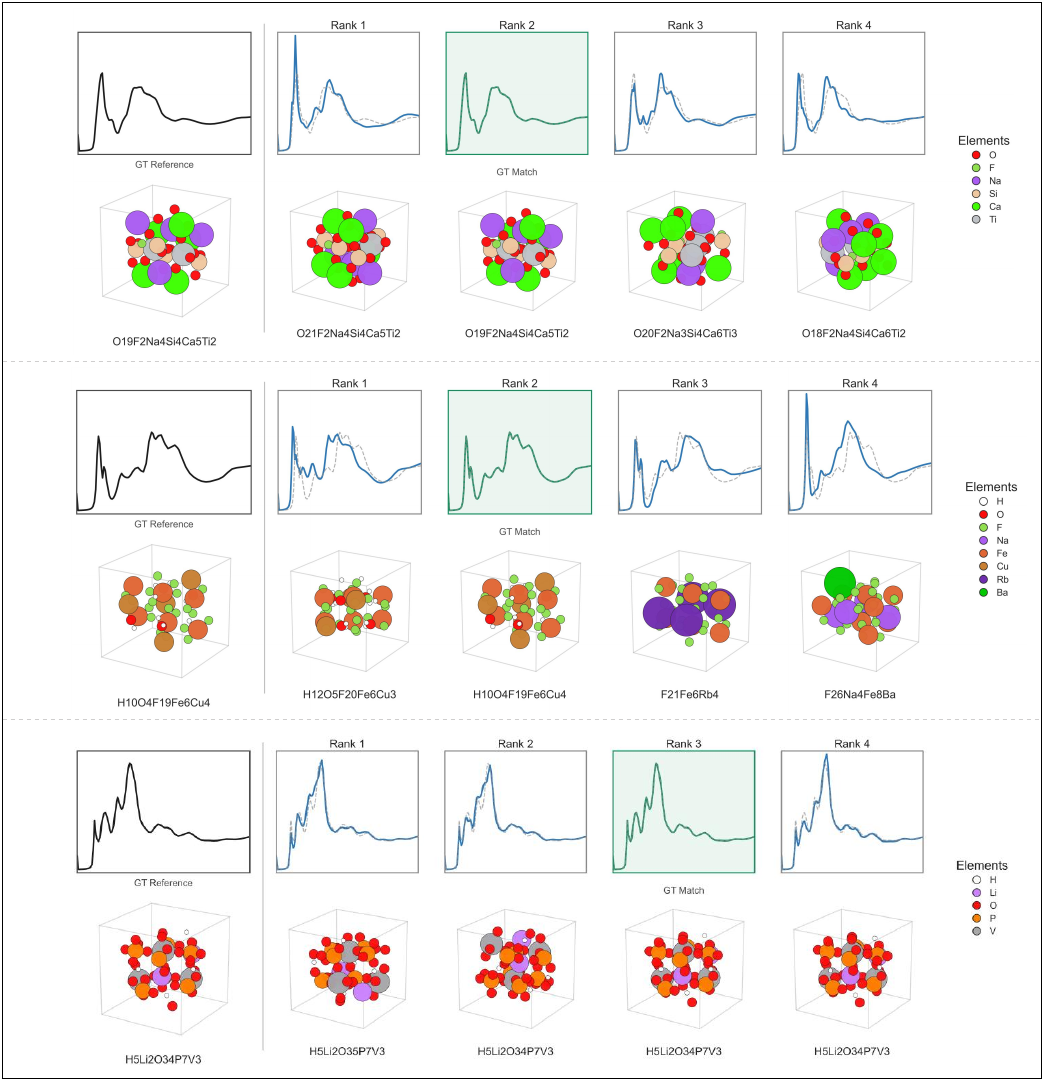}
    \caption{Extended qualitative examples of spectrum-to-structure retrieval across diverse absorbing elements. For each case, the input is a query spectrum (top-left).
    The remaining panels in the top row show the spectra associated with the retrieved candidate structures, ranked by similarity in the latent space. The bottom row visualizes the corresponding local structures: the ground-truth structure of the query (bottom-left) and the retrieved candidate structures (right). Green bounding boxes indicate the correct match among the ranked results. Note that the spectra in the top row (except the query) are not model predictions, but are shown for visualization to illustrate spectral–structural consistency. These examples demonstrate that the model retrieves either the exact coordination environment or physically highly similar local structures, reflecting effective cross-modal alignment.}
    \label{fig:retrieval}
\end{figure*}

\begin{figure*}[htbp]
    \centering
     \includegraphics[width=0.9\linewidth]{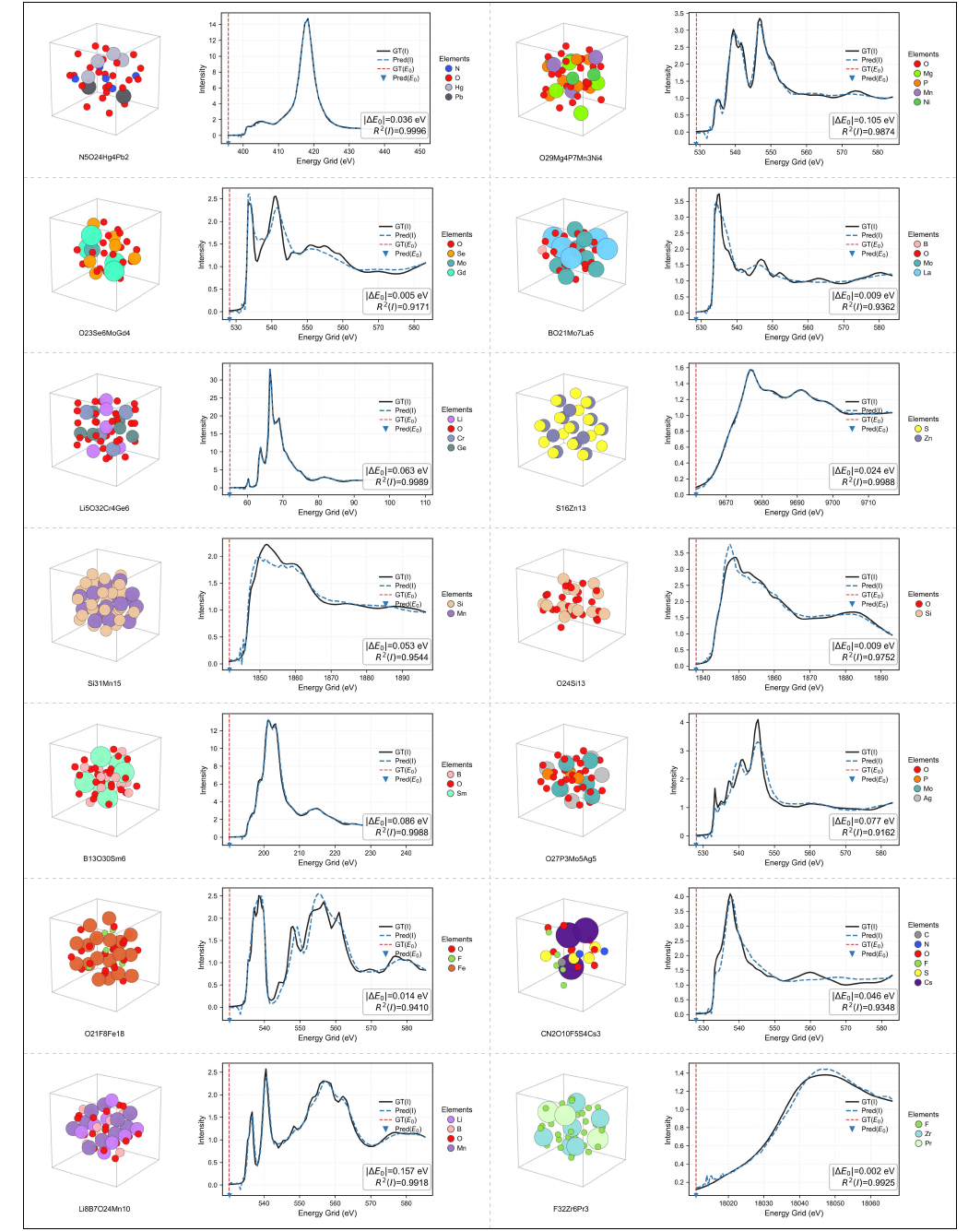}
        \caption{Extended qualitative examples of forward anchored absolute-spectrum prediction. Each panel displays the input absorber-centered 3D local structure alongside the corresponding ground-truth (GT) and predicted XANES spectra. The inset metrics denote the absolute energy anchor error ($|\Delta E_0|$ in eV) and the coefficient of determination ($R^2$) for the intensity profile, highlighting the model's ability to preserve absolute physical scales and fine-grained spectral morphology.}
    \label{fig:forward}
\end{figure*}
    
\begin{figure*}[htbp]
    \centering
    \includegraphics[width=0.9\linewidth]{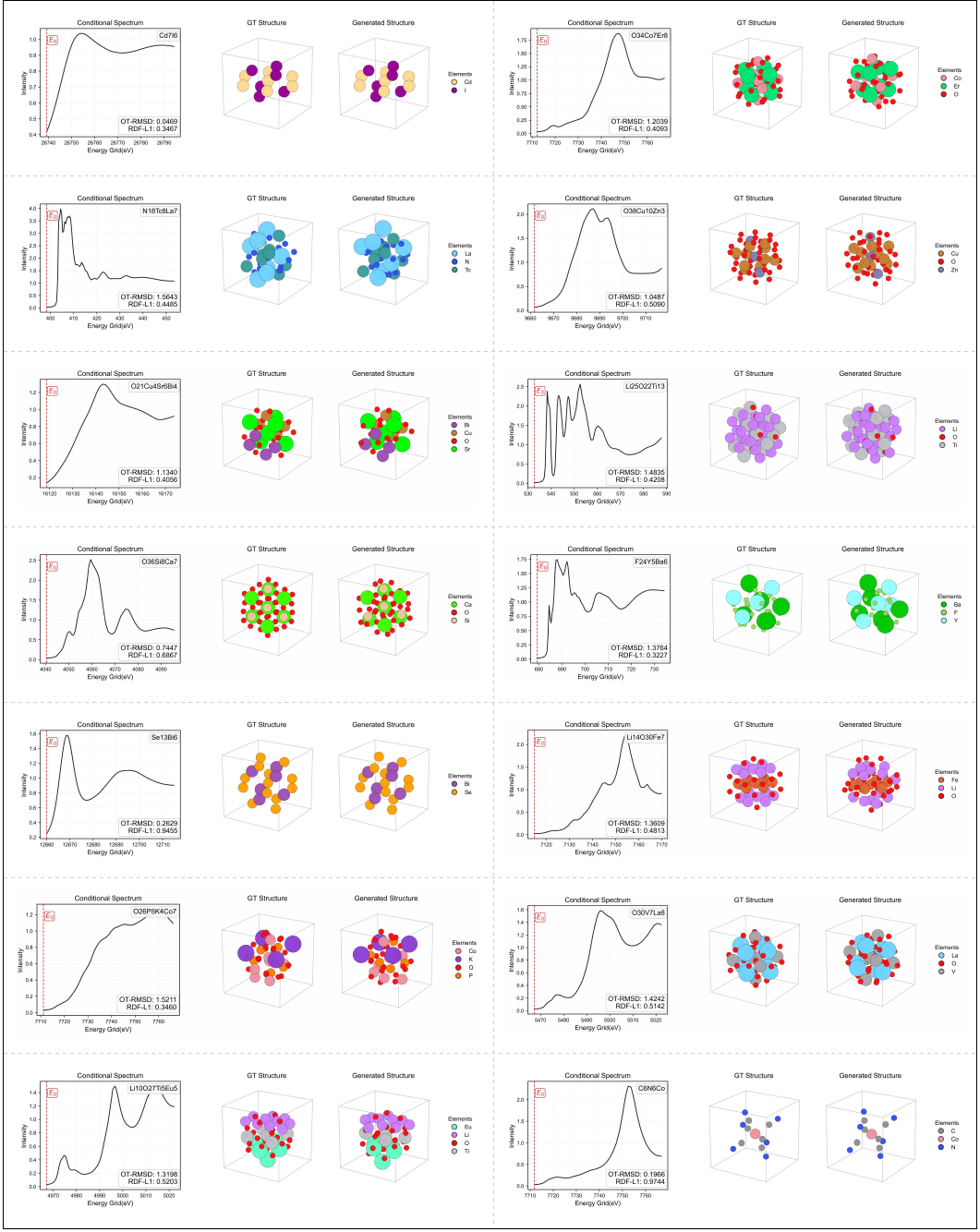}
        \caption{Extended qualitative examples of inverse composition-conditional 3D structure generation. Each panel presents the input 1D spectral condition (left) and compares the GT local atomic environment against the 3D structure generated by Uni-XAS (right). The visualizations, together with the annotated OT-RMSD and RDF-L1 metrics, provide representative examples of the model's generated multi-shell geometries under the benchmark setting used in this work.}
        \label{fig:inverse}
\end{figure*}

\section{Extended Qualitative Visualizations}
\label{app:visualizations}

    To keep the organization of this supplement aligned with the main paper, the qualitative figures are placed in this section and should be read in conjunction with Figure~3, Table~2, and Table~3 of the main text.
    
    \paragraph{XASLip Cross-Modal Retrieval.}
    Figure~\ref{fig:retrieval} displays extended qualitative examples of the spectrum-to-structure retrieval task. Each case is conditioned on an input spectrum, and the model retrieves candidate local structures ranked by similarity in the shared latent space. These examples illustrate the alignment between spectral features and structural coordination across diverse absorbing elements. The visualizations demonstrate that the model is able to retrieve the exact ground-truth coordination environment (highlighted in green) or highly similar local structures given an input spectral reference.
    
    \paragraph{Forward Absolute-Spectrum Prediction.}
    Figure~\ref{fig:forward} presents held-out examples exported under a deterministic selection protocol rather than manual cherry-picking. These examples complement Table~2 of the main text by showing that the decomposed physical prediction head and retrieval-aware consistency constraint jointly preserve both the absolute edge anchor and the fine-grained near-edge morphology across diverse absorbers.
    
    \paragraph{Inverse Local-Structure Generation.}
    Figure~\ref{fig:inverse} presents held-out spectrum-conditioned local environments evaluated with the same OT-RMSD and RDF-L1 metrics used in Table~3 of the main text. Together with the quantitative results, these visualizations clarify what the PR-Flow generator recovers reliably, and where residual geometric ambiguity remains under the one-to-many inverse mapping.

\end{document}